%% file: switchable_main.tex
\newif\ifjcp
\newif\ifacs
\newif\ifrsc

\jcptrue
\newcommand{\papertitle}{Low-rank compression of two-electron reduced density matrices}

\newcommand{\setinfo}{%
    \title{\papertitle}%
    \author{Kemal Atalar}%
    \email{kemal.atalar@kcl.ac.uk}
    \affiliation{Department of Physics and Thomas Young Centre, King’s College London, Strand, London, WC2R 2LS, UK}%
    \author{Hugh G. A. Burton}
    \affiliation{Department of Chemistry, University College London, London, WC1H 0AJ, UK}
    \author{Andreas Gr\"uneis}
    \affiliation{Institute for Theoretical Physics, TU Wien, Wiedner Hauptstraße 8-10/136, Vienna, Austria}
    \author{George H. Booth}%
    \email{george.booth@kcl.ac.uk}%
    \affiliation{Department of Physics and Thomas Young Centre, King’s College London, Strand, London, WC2R 2LS, UK}%
}

\input{preamble}

\ifacs
    \setinfo
\fi
\begin{document}
\ifjcp
    \setinfo
\fi

\ifrsc
\else
    \begin{abstract}
    \input{abstract}
    \end{abstract}
\fi

\ifjcp
    \maketitle
\fi

%\tableofcontents

%%%%%%%%%%%%%%%%%%%%%%%%%%%%%%%%%%%
% Body of the document goes here: %
%%%%%%%%%%%%%%%%%%%%%%%%%%%%%%%%%%%

\input{1-introduction}

%\input{2-methodology}

\input{3-results1}

\input{4-results2}

\input{5-results-evcont}

\input{6-conclusion}
\ifacs
    \newpage
\fi

% Acknowlegements:
\ifrsc
    \section*{Conflicts of interest}
    There are no conflicts to declare.
\fi

\input{acknowledgements}

\section*{References}
% Set the bibliography:
\ifjcp
    \bibliographystyle{aipnum4-2}
    \bibliography{ref_submission}
\else
    \ifrsc
        \bibliography{ref}
        \bibliographystyle{rsc}
    \else
        \bibliography{ref.bib}
    \fi
\fi

% Supporting information:
\input{appendix}

%\newif\ifsi
%\sifalse  % Comment/uncomment to switch

% Define style:
%\ifsi
% Supporting information:
%\clearpage
%\onecolumngrid
%\section*{Supporting Information}

%\subsection*{Input for Chlorine dimer results of Fig.~\ref{fig:cl2_intro}, providing all energy functionals described}
%\inputminted[
%bgcolor=LightGray,
%fontsize=\footnotesize,
%]{python}{demo_scripts/cl2.py}

%\twocolumngrid

%\clearpage
%\section*{Previous Figures}
%\input{old_figures}
%\clearpage

%\section*{Previous Text}
%\input{old_text}
%\clearpage

% End the document:
\end{document}

%% file: preamble.tex
\ifjcp
    \documentclass[twocolumn,aip,jcp,reprint,floatfix]{revtex4-2}
\fi

\ifacs
    \documentclass[journal=jacsat,manuscript=article]{achemso}
    \setkeys{acs}{articletitle=true}
    \newcommand{\onlinecite}[1]{\hspace{-1ex}\nocite{#1}\citenum{#1}}
\fi

\ifrsc
    \documentclass[twoside,twocolumn,10pt]{article}
    \usepackage{extsizes}
    \usepackage[super,sort&compress,comma]{natbib}
    \usepackage[version=3]{mhchem}
    \usepackage[left=2cm,right=2cm,top=1.785cm,bottom=2.0cm]{geometry}
    \usepackage{balance}
    \usepackage{mathptmx}
    \usepackage{sectsty}
    \usepackage{lastpage}
    \usepackage[format=plain,justification=justified,singlelinecheck=false,
        font={stretch=1.125,small,sf},labelfont=bf,labelsep=space]{caption}
    \usepackage{fancyhdr}
    \usepackage{fnpos}
    \usepackage[english]{babel}
    \addto{\captionsenglish}{}
    \usepackage{array}
    \usepackage{droidsans}
    \usepackage{charter}
    \usepackage[T1]{fontenc}
    \usepackage{setspace}
    \usepackage[compact]{titlesec}
    \usepackage{epstopdf}
\fi

\usepackage{float}
\usepackage[usenames,dvipsnames]{xcolor}
\usepackage{graphicx}
\usepackage{dcolumn}
\usepackage{bm}
\usepackage{amsmath}
\usepackage{braket}
\usepackage{dsfont}
\usepackage{hyperref}
\newif\ifuseminted
\usemintedfalse
\ifuseminted
    \usepackage{minted}
\else
    \usepackage{listings}
    
\fi
\usepackage{nicefrac}
\usepackage{amssymb}
\usepackage{lipsum}

\newcommand{\toremove}[1]{\ignorespaces}

\makeatletter
\newcommand{\ifemptyarg}[3]{%
    \if\relax\detokenize\expandafter{#1}\relax
        #2%
    \else
        #3%
    \fi
}
\newcommand{\ifstringequal}[4]{%
    \edef\@tempA{\detokenize\expandafter{#1}}%
    \edef\@tempB{\detokenize\expandafter{#2}}%
    \ifx\@tempA\@tempB #3\else #4\fi
}
\newcommand{\appendemail}[2]{%
    \ifemptyarg{#2}{}{%
        \ifx#1\@empty
            \edef#1{#2}%
        \else
            \edef#1{#1; #2}%
        \fi
    }%
}

\makeatother

\ifjcp
\else
\fi

\ifrsc
    \fancypagestyle{plain}{
        
    }
    \makeFNbottom
    \makeatletter
    \renewcommand\LARGE{\@setfontsize\LARGE{15pt}{17}}
    \renewcommand\Large{\@setfontsize\Large{14pt}{14}}
    \renewcommand\large{\@setfontsize\large{12pt}{12}}
    \renewcommand\footnotesize{\@setfontsize\footnotesize{7pt}{10}}
    \makeatother

    \makeatletter
    \renewcommand\@biblabel[1]{#1}
    \renewcommand\@makefntext[1]{\noindent\makebox[0pt][r]{\@thefnmark\,}#1}
    \makeatother
    
    \sectionfont{\sffamily\Large}
    \subsectionfont{\normalsize}
    \subsubsectionfont{\bf}
    \titlespacing*{\section}{0pt}{4pt}{4pt}
    \titlespacing*{\subsection}{0pt}{15pt}{1pt}
    \makeatletter
    \newlength{\figrulesep}
    \makeatother
\fi

%% file: abstract.tex
Two-body reduced density matrices (2RDMs) encode the essential two-electron physics of electronic states, but their quartic storage cost poses a major limitation in practical workflows. We investigate a simple protocol to compress both transition and non-transition 2RDMs into a lower-rank representation that preserves their wedge-product structure and physical symmetries under truncation. The resulting decomposition couples Coulomb and exchange channels through a common set of low-rank factors, yielding a more compact rank-sparse representation than single-channel factorizations. For correlated states, the effective rank scales linearly with system size, achieving a $\sim99$\% compression for the coupled-cluster 2RDM of octane while retaining chemical accuracy. We apply this to the recently introduced {\em ab initio} eigenvector continuation workflows, where many-body wave functions are interpolated across nuclear geometries with mean-field cost. Here, 2RDMs between training states act as projectors into a subspace but their memory scaling limits applications to larger systems. The compression scheme reduces the memory cost from quartic to quadratic for a fixed error per electron. Metrics to systematically control the decomposition are investigated, enabling statistically resolved structural, dynamical and spectroscopic observables from nonadiabatic molecular dynamics simulations of photoexcited H$_{28}$ chains, interpolating from compressed near-exact DMRG training data. This establishes these structure-preserving compressed intermediates for practical correlated electronic structure workflows. 
%\ka{I think it's worth mentioning that the protocol optimally compresses formally low-rank RDMs such as those obtained from complete-active space states.} \ghb{I think unless we find things to remove, we shouldn't add more stuff now...}
%A truncation threshold of 1~mHa is found to recover structural, dynamical and spectroscopic observables quantitatively on the timescales relevant to the dynamics. 
%This establishes a low-rank, structure-preserving two-body variable as a practical compressed intermediate for correlated electronic structure workflows.

%% file: 1-introduction.tex
\section{Introduction}

The two-body reduced density matrix (2RDM) is a central object in the description of interacting many-electron systems, providing a compact representation of the pair correlations present in a many-body wavefunction. In particular, given the 2RDM, expectation values of all static two-electron operators can be evaluated, so that it encodes the essential two-body physics of the state. Because the electronic Hamiltonian contains at most two-body interactions, this also permits the computation of the in-principle exact energy of an electronic state, making the 2RDM a particularly important compressed quantum variable for correlated many-body problems. More generally, pair correlations \emph{between} distinct many-body states are encoded in the \emph{transition} two-body reduced density matrix (2tRDM), which governs transition matrix elements and therefore characterizes, among other quantities, couplings, excitation character, and transport-related observables.

The 2RDM is not an arbitrary four-index object, but is subject to strong algebraic and physical constraints that endow it with substantial structure~\cite{doi:10.1021/cr2000493,https://doi.org/10.1002/qua.560200604,Kutzelnigg10022010}. For an uncorrelated single-determinant state, the 2RDM is obtained exactly as the wedge product of the corresponding 1RDM, explicitly revealing the low-rank structure inherent to mean-field descriptions. Irreducible two-body correlations beyond this factorized form are captured by a size-extensive cumulant contribution, which measures departures from the 1RDM product structure and generally increases the effective rank with growing system size~\cite{10.1063/1.478189}.
%Beyond these structures, $N$-representability conditions more generally constrain the 2RDM to ensure its derivability from a valid wavefunction ansatz.

Understanding and exploiting this low-rank structure in 2(t)RDMs is therefore of considerable interest. If the correlation-induced part of the 2RDM can be represented compactly, then the storage, communication, manipulation of two-body information and evaluation of expectation values may be significantly accelerated in correlated electronic-structure methods, or effectively learnt in machine-learning workflows~\cite{b2026machinelearningtwoelectronreduced}. At the same time, such a representation offers a natural way to quantify the complexity of irreducible two-body physics in a state or approximate wavefunction, e.g. in the characterization of extensive coherent quantum phenomena~\cite{PhysRevA.92.052502,10.1063/1.3256237}. More broadly, low-rank tensor factorizations of two-body quantities have long been explored in electronic-structure theory, both to reduce computational cost and to expose physically meaningful fluctuation spaces. In particular, decompositions of suitably partitioned two-body objects can define compact fluctuation manifolds that exploit emergent locality of correlation and enable reduced-scaling approaches~\cite{PhysRevA.67.052504,Mazziotti12-lowrank,Kinoshita03-SVD-CCSD,HeadGordon04-SVD-T2,Bell10-HOSVD-MP2,Benedikt11-CP-decomp-MP2,Yang12-virtualorbopt-CCSD,PhysRevB.104.245114,10.1063/1.5092505,Martinez12-THC-T2,PhysRevB.102.165107}.

Rank-revealing decompositions of the 2RDM itself have been investigated especially in variational reduced-density-matrix methods, in which the 2RDM is optimized directly under approximate enforcement of $N$-representability constraints~\cite{Mazziotti06-v2RDM,Mazziotti07-parametric2RDM,https://doi.org/10.1002/wcms.1702}. In that setting, low-rank approximations have been shown to offer substantial computational advantages by exposing intrinsic structure in the 2RDM and reducing the formal scaling of the method~\cite{Mazziotti07-lowrank-v2RDM,Mazziotti12-lowrank,Mazziotti13-comparison_lowrank}. More recently, rank-exposed representations of reduced density matrices have attracted renewed attention in the context of matrix completion and noise filtering for incomplete 2RDM data, motivated in part by quantum algorithms in which these quantities must be estimated from measurements~\cite{doi:10.1021/acs.jpclett.6c00296,Chan23-RDM-completion,anselmetti2025classicaloptimisationreduceddensity}. In Ref.~\onlinecite{Chan23-RDM-completion}, for example, low-rank 2RDM representations were shown to improve resilience to statistical noise by filtering incomplete or noisy measurement data.

% where restricting the rank of the variational optimization was found to be analogous to an active-space approximation in multiconfigurational self-consistent field methods.\cite{Mazziotti07-lowrank-v2RDM} Similarly, representing the 2RDM through truncated spectral decompositions in parametric formulations produced speedups of several orders of magnitude.\cite{Mazziotti12-lowrank} Comparisons between different low-rank forms have further indicated that spectral decompositions can capture the intrinsic structure of 2RDMs more efficiently than higher-order tensor contractions such as tensor hypercontraction.\cite{Mazziotti13-comparison_lowrank}

%Beyond classical electronic structure methods, low-rank structure in reduced density matrices is also relevant in emerging quantum algorithms. For example, Peng \textit{et al.}\cite{Chan23-RDM-completion} demonstrated that low-rank approximations can be used for both noise filtering and matrix completion of sampled 2RDMs, improving the reconstruction of two-body observables from incomplete or noisy measurements. \ka{Should I comb for noise filtering and matrix completion references from Chan paper to add here?}

In this work, we investigate a general, simple, and robust decomposition that exposes the intrinsic low-rank structure of both 2RDMs and transition 2RDMs. Unlike more generic matrix factorizations, the present decomposition is constructed to preserve fermionic antisymmetry and the wedge-product structure of the two-body density matrix under arbitrary truncation. As a result, the compressed representation remains physically consistent even at reduced rank. In Sec.~\ref{sec:scaling_inference}, we then investigate the systematic truncation of these two-body correlations to obtain compressed 2(t)RDM representations across a range of electronic-structure methods and correlation regimes, including full configuration interaction (FCI), complete active space self-consistent field (CASSCF), coupled cluster (CCSD)~\cite{RevModPhys.79.291}, and matrix product states (MPS)~\cite{10.1063/1.2883980}. We compare this compression to alternative low-rank approximations, analyze its scaling with system size for alkane chains, and derive compact expressions for evaluating two-electron observables and nuclear derivatives directly from the compressed representation.

Our particular motivation for developing such a low-rank expansion comes from the recently developed \emph{ab initio} eigenvector continuation (EC) framework~\cite{RevModPhys.96.031002} for interpolating correlated many-body wavefunctions across molecular geometries~\cite{mejutozaera2023-evcont,Rath25-evcont}. In this framework, a set of training wavefunctions is computed at selected geometries using an accurate electronic-structure solver, and the electronic problem at a new geometry is projected into the subspace spanned by these states to define a low-dimensional effective Hamiltonian. Transition two-body reduced density matrices between training states then provide the matrix elements required to project the electronic Hamiltonian at arbitrary geometries into this reduced subspace. An important feature of this construction is that the cost of the projection step, once the 2(t)RDMs are available, is independent of the complexity of the many-body solver used to generate the training states. This enables the use of highly accurate solvers, including density matrix renormalization group (DMRG) methods, in the construction of the effective Hamiltonian, and has recently opened the door to applications such as correlated nonadiabatic molecular dynamics~\cite{Rath25-evcont,Atalar24-eigconNAMD-faraday}. However, the quartic memory scaling associated with storing the full set of 2tRDMs becomes a significant bottleneck as the system size, basis size, and number of training states increase. Efficient compression of these objects is therefore essential for extending EC-based electronic-structure interpolation to larger and more realistic applications.

In Sec.~\ref{sec:NucEnsem}, we combine the proposed rank-compression scheme with eigenvector continuation to demonstrate the practical utility of compressed 2(t)RDM representations within an interpolation workflow for correlated electronic structure. Using compressed MPS-based representations across nuclear ensembles, we quantify the effect of the compression on inferred energies and on the sampling of nuclear phase space in nonadiabatic molecular dynamics simulations. This represents an important step towards efficient and practical use of highly-correlated methods across molecular dynamics timescales.

%In Sec.~\ref{sec:NucEnsem}, we therefore put this rank-compression scheme together with the eigenvector continuation interpolation of quantum states, to demonstrate the utility of these compressed representations within an interpolation workflow for correlated electronic structure. Applying compressed MPS representations across nuclear ensembles, we quantify the impact of the compression on inferred energies of the model, and on the sampling of nuclear phase space in nonadiabatic molecular dynamics simulations.

%% file: 3-results1.tex
\section{Spectral decomposition of 2RDMs} \label{sec:decomp}

We define the two-body (transition) reduced density matrix (2tRDM) between many-body states $|\Psi_a\rangle$ and $|\Psi_b\rangle$ as
\begin{equation}
{}^{ab}\Gamma_{ijkl}
=
\sum_{\sigma, \tau} \langle \Psi_a |
\hat c_{i \sigma}^\dagger \hat c_{k \tau}^\dagger \hat c_{l \tau} \hat c_{j \sigma}
| \Psi_b \rangle,
\label{eq:2trdm-definition}
\end{equation}
where $\hat c^\dagger$ and $\hat c$ denote fermionic creation and annihilation operators in an orthonormal basis, and $\sigma, \tau$ are spin variables. Throughout this work we employ a spin-free formulation in which spin degrees of freedom have been summed over in the operators and spin is integrated out of the expressions, while we will also assume real-valued states~\cite{Kutzelnigg10022010}. This allows the electronic energy of a state, $|\Psi_a\rangle$, to be given by
\begin{equation}
    E_a = \mathrm{Tr}[K_{ijkl} {}^{aa}\Gamma_{ijkl}] ,
\end{equation}
where $K_{ijkl}$ is the reduced Hamiltonian, given by
\begin{equation}
    K_{ijkl} = \frac{1}{N-1} h_{ij} \delta_{kl} + \frac{1}{2}(ij|kl) , \label{eq:energyfunc}
\end{equation}
%\ghb{Happy with this expression?} \ka{Mazziotti defines this such that both 1RDM traces equally contribute, i.e. $0.5*(h_{ij} \delta_{kl} + h_{kl} \delta_{ij})$. But I am guessing if 1RDM from both traces is identical, this expression is true as well. It's only a problem if 2RDM gives different partial traces.} \ghb{Can our truncated low rank form give different 1RDM traces depending on the electron which is traced out?}
with $N$ the number of electrons, $h_{ij}$ the one-body part of the Hamiltonian, and $(ij|kl)$ the two-electron repulsion integrals in Mulliken notation. This form implicitly uses the fact that the 1(t)RDM can be found by integrating out an electron coordinate from the 2(t)RDM, as
\begin{align}
    {}^{ab}\gamma_{ij} &= \sum_{\sigma} \langle \Psi_a |
\hat c_{i \sigma}^\dagger \hat c_{j \sigma}
| \Psi_b \rangle \\
    &= \frac{1}{N-1}\sum_k {}^{ab}\Gamma_{ijkk} . \label{eq:1trdm}
\end{align}
An important property for the spin-summed 2(t)RDM is the symmetry under electron exchange and adjoint relation, which holds for both the transition and non-transition cases, as 
\begin{equation}
    {}^{ab}\Gamma_{ijkl}={}^{ab}\Gamma_{klij}=\left({}^{ba}\Gamma_{jilk} \right)^* = \left({}^{ba}\Gamma_{lkji} \right)^*.
\end{equation}

For single-determinant states, the 2tRDM is given by the (overlap weighted) wedge product of the corresponding one-body transition density matrices,
\begin{equation}
{}^{ab}\Gamma_{ijkl}
=
\frac{1}{{}^{ab}S}
\,{}^{ab}\gamma_{ij}\,
{}^{ab}\gamma_{kl}
-
\frac{1}{2\,{}^{ab}S}
\,{}^{ab}\gamma_{il}\,
{}^{ab}\gamma_{kj},
\label{eq:hf-2rdm}
\end{equation}
where ${}^{ab}S=\langle \Psi_a | \Psi_b\rangle$ denotes the (non-zero) overlap between the states~\cite{10.1063/5.0045442,10.1063/5.0122094} (we generalize to the case of ${}^{ab}S=0$ in Sec.~\ref{sec:tRDM}). This provides a low-rank structure of the 2tRDM which is independent of system size in mean-field methods. The two-electron terms which are generated in the energy functional of Eq.~\ref{eq:energyfunc} are denoted the Coulomb (first term of Eq.~\ref{eq:hf-2rdm}) and exchange (second term) energies respectively.
Rank-increasing deviations from the form of Eq.~\ref{eq:hf-2rdm} therefore directly reflect the presence of irreducible two-body correlations in the cumulant contribution~\cite{https://doi.org/10.1002/qua.997}. 

\subsection{Single channel decompositions} \label{sec:singlechanneldecomps}

We first focus on the non-transition `same-state' 2RDM between the same state, dropping the state indices ($a,b$) for simplicity, and generalizing to transition RDMs in Sec.~\ref{sec:tRDM}. The simplest way to expose low-rank structure of the 2RDM in Eq.~\ref{eq:2trdm-definition} is to reshape the four-index tensor into a matrix and spectrally decompose it in an appropriate pair space of natural geminal functions~\cite{https://doi.org/10.1002/qua.560020203}. However, this requires a choice of how to group the four orbital indices into two compound indices. This matricization choice is not unique, and different groupings emphasize different physical structures.
The structure of the 2RDM over mean-field states in Eq.~\ref{eq:hf-2rdm} suggests either a Coulomb or exchange index pairing. In the Coulomb grouping, the rank is exposed between the pairs of indices corresponding to electron one, $(ij)$, and electron two, $(kl)$ in Eq.~\ref{eq:2trdm-definition}, aligning with the direct product structure $\gamma_{ij}\gamma_{kl}$ that generates Coulomb (Hartree) energy contributions. This leads to the spectral decomposition
\begin{equation}
\Gamma_{ij,kl}
= \sum^{~r}_{\alpha} v_{ij}^{(\alpha)}\, \epsilon^{(\alpha)}\, v_{kl}^{(\alpha)},
\label{eq:coulomb_grouping}
\end{equation}
where the composite index $(ij)$ labels two orbitals represented by the low-rank vectors of orthonormal pair functions, $v_{ij}^{(\alpha)}$, with pair occupations $\epsilon^{(\alpha)}$. The rank $r$, has a maximum value at full rank of the square of number of orbitals, $\mathcal{O}[M^2]$. Truncating this expansion at $r$ values less than the full rank yields a low-rank approximation, where correlations between instantaneous fluctuations in electron densities of the states are approximated.

Alternatively, one may permute the $j \leftrightarrow l$ indices of the density matrix and group the indices as $(il)$ and $(kj)$, yielding a symmetric matrix with the rank-exposed form
\begin{equation}
{\tilde \Gamma}_{il,kj}
= \sum^{~r}_{\alpha} w_{il}^{(\alpha)}\, \lambda^{(\alpha)}\, w_{kj}^{(\alpha)}.
\label{eq:exchange_grouping}
\end{equation} 
%\ka{It is symmetric: $\Gamma_{ijkl} \neq \Gamma_{ilkj}$ BUT $\Gamma_{ilkj} = \Gamma_{kjil}$ for real diagonal RDMs! ($<i^\dagger k^\dagger l j> = <k^\dagger i^\dagger j l>$)  That's where the confusion comes. So, this permutation is symmetric wrt pair swap, just not symmetric wrt original Gamma.} \ka{This also explains why Q is guaranteed to be symmetric because both the Coulomb and exchange transpositions are symmetric.}
This pairing creates natural pair geminals which connect a creation operator with an annihilation on the different particle line and therefore reflects the permutation structure $\gamma_{il}\gamma_{kj}$ associated with generating exchange (Fock) interaction energies from the antisymmetry of the state. 
There is a third inequivalent matricization, spectrally decomposing the object ${\hat \Gamma}_{ik,lj}$ which groups the creation and annihilation operators together, exposing the geminals characterizing instantaneous electron number fluctuations through the particle-particle channel. We refer to this decomposition as the `cross' grouping, which was the choice utilized in the work of Peng. et al.\cite{Chan23-RDM-completion}.

All three matricizations are exact rearrangements of the same four-index tensor and therefore contain the same information at full rank. However, they define different matrix unfoldings of the 2RDM, associated with different physical channels and with the $D$, $Q$, and $G$ metrics familiar from reduced density matrix theory~\cite{PhysRevA.71.062503}. As a consequence, their singular or spectral profiles, and therefore their optimal low-rank truncations, generally differ. Any one such unfolding treats the 2RDM as a simple tensor product object in a single channel, and therefore does not by itself preserve the antisymmetrized wedge-product structure that characterizes fermionic two-particle densities. The latter is instead encoded jointly through the permutation symmetries of the 2RDM, as well as the cumulant contribution that captures irreducible two-body correlations beyond the mean-field wedge product of Eq.~\ref{eq:hf-2rdm}~\cite{https://doi.org/10.1002/qua.997}.

\subsection{Joint decomposition}
\label{sec:joint}

Decompositions based on a single matricization therefore do not, in general, expose the most compact low-rank structure of fermionic two-body densities. This limitation was already observed in variational parametric 2RDM methods, where simultaneous parameterization of multiple channels was found to outperform any single-channel expansion~\cite{Mazziotti12-lowrank}. The underlying reason is simple: even formally low-rank 2RDMs, such as those arising from mean-field or fixed active-space states, are not naturally low-rank in any one channel alone, because the fermionic wedge-product structure couples direct and exchange channels.
%Decompositions of 2RDMs using a single index grouping therefore do not naturally expose the appropriate low-rank structure, and as a consequence are not generally compact. This was observed previously in variational parameteric 2RDM methods, where it was shown that parameterizing contributions simultaneously from the three different spectral expansions vastly outperformed any single expansion~\cite{Mazziotti12-lowrank}. 

\begin{figure}[tb]
\centering
\includegraphics[width=0.98\linewidth]{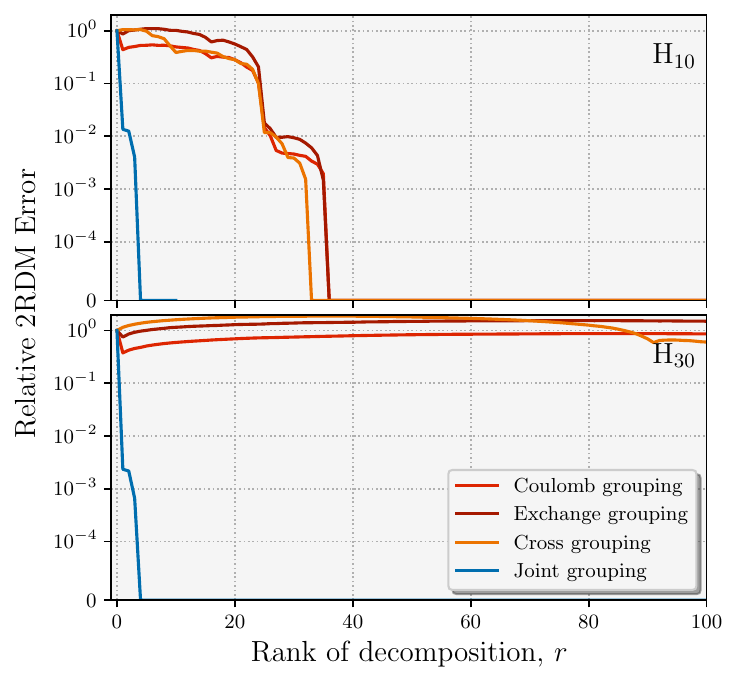}
  \caption{Low-rank decomposition of CAS(2,2)SCF two-body reduced density matrices for linear H$_{10}$ (top) and H$_{30}$ (bottom) chains in a minimal basis at 1.5a$_0$ internuclear separation under different spectral decompositions.}
  %\includegraphics[width=0.95\linewidth]{Figs/H30_cas(2,2)scf_lr_decompositions_normonly.pdf}
  %\caption{Low-rank decomposition of CAS(2,2)SCF two-body reduced density matrix of linear H$_{30}$ chain in minimal basis using different grouping of orbital indices.}
  \label{fig:cas-decomp}
\end{figure}

We can demonstrate this simply by considering a small CAS with 2 electrons in 2 orbitals -- CAS(2,2) -- each state consists of at most four Slater determinants. As a consequence, the complete 2RDM decomposition is bounded by the finite dimension of the active-space determinant manifold, which should not scale with system size, and involves a rank of at most only ten unique cross-determinant contributions. This should bound the rank of any 2(t)RDM decomposition, regardless of system size.
Figure~\ref{fig:cas-decomp} illustrates low-rank truncations via the pair occupation numbers in the decomposition into Coulomb, exchange and cross groupings. For linear chains of 10 and 30 hydrogen atoms, it shows the relative error over the 2RDM elements from rank decompositions of a CAS(2,2) density matrix. The number of vectors required to accurately reconstruct the 2RDM substantially exceeds the formal complete rank limit of a CAS(2,2) state, and furthermore scales with system size, demonstrating the challenge of these individual groupings. %These results clearly demonstrate that decompositions into individual groupings cannot fully exploit the low-rank structure of the 2RDM and therefore require more vectors than the formal minimum to achieve the same accuracy.

This suggests seeking a `joint' decomposition in which the direct and exchange channels are not truncated independently, but are constrained to arise from the same underlying set of factors, thereby preserving the wedge-product structure characteristic of fermionic two-body densities, as
\begin{equation}
    \Gamma_{ijkl} = \sum^{~r}_{\alpha} \varepsilon^{(\alpha)}\big[ v_{ij}^{(\alpha)} v_{kl}^{(\alpha)} \, - \, \frac{1}{2} v_{il}^{(\alpha)} v_{kj}^{(\alpha)} \big].
    \label{eq:joint_decomp}
\end{equation}
%where the eigenvalues of the spectral decomposition can be absorbed into the spin-free pair-generating factors, $v_\alpha$. 
This can effectively represent the 2RDM as a sum over contributions analogous to Slater determinant pairs, ensuring the appropriate low-rank CAS form can be found. This also treats both electrons in the 2RDM consistently, ensuring that the same 1RDM is found regardless of which electron is traced out from Eq.~\ref{eq:1trdm}.

To obtain these vectors, $v^{(\alpha)}$, in the joint grouping, one can re-label the $j$ and $l$ indices of Eq.~\ref{eq:joint_decomp} as:
\begin{equation}
    \Gamma_{ilkj} = \sum^{~r}_{\alpha} \varepsilon^{(\alpha)}\big[ v_{il}^{(\alpha)} v_{kj}^{(\alpha)} \, - \, \frac{1}{2} v_{ij}^{(\alpha)} v_{kl}^{(\alpha)} \big].
    \label{eq:joint_decomp_relabel}
\end{equation}
We can linearly combine the previous two expressions in order to algebraically cancel the exchange contribution and identify an auxiliary two-body variable, $Q_{ijkl}$, which is symmetric and has the desired outer product form within the Coulomb grouping. This can be spectrally decomposed as
\begin{equation}
Q_{ijkl} = \frac{4}{3}\Gamma_{ijkl} + \frac{2}{3}\Gamma_{ilkj} =
\sum^{~r}_{\alpha} v_{ij}^{(\alpha)} \varepsilon^{(\alpha)} \, v_{kl}^{(\alpha)}
\label{eq:coulombvec_jointdecomp}
\end{equation}
to directly obtain the pair vectors, $v_{ij}^{\alpha}$, that contribute via a wedge product form to both Coulomb and exchange channels as in Eq.~\ref{eq:joint_decomp}. These $v_{ij}^{\alpha}$ vectors cannot be considered geminals in the traditional sense, and are not generally symmetric or antisymmetric with respect to exchange of their spatial indices, but do form an orthonormal set. Furthermore, since $Q_{ijkl}$ is not positive semi-definite (albeit Hermitian), the eigenvalues, $\varepsilon^{(\alpha)}$, can not be considered as geminal occupations, and truncations to low-rank expansions should be performed based on their absolute value.
While spectral decomposition properties of 2RDMs and truncations of natural geminal expansions have long been discussed in the literature, with low-rank parameterizations explored in variational 2RDM theory, we are not aware of a broadly adopted low-rank electronic structure framework based on the wedge-preserving coupled decomposition of Eq.~\ref{eq:coulombvec_jointdecomp}~\cite{10.1063/1.1669767,https://doi.org/10.1002/qua.560020203,Kutzelnigg10022010,10.1063/1.3256237,https://doi.org/10.1002/qua.560200604,Mazziotti12-lowrank,doi:10.1021/cr2000493,PhysRevA.71.062503}.

By construction, this `joint' decomposition is able to find the appropriate formal low-rank decomposition of the 2RDMs from mean-field (strictly one non-zero eigenvector of $Q_{ijkl}$) and CAS states. Figure~\ref{fig:cas-decomp} shows how this decomposition exactly describes the CAS(2,2)SCF 2RDMs with only $r=4$ vectors, guaranteeing maximum efficiency in the compression, with the rank only determined by the irreducible two-body physics of the fixed-size active space. The compression to 4 vectors is achieved for both H$_{10}$ and H$_{30}$ cases in Fig.~\ref{fig:cas-decomp} in contrast to the decomposition into single groupings. 

\subsection{Diagonal contributions}
\label{sec:diag_corr}

\begin{figure}[tb]
\centering
  \includegraphics[width=0.95\linewidth]{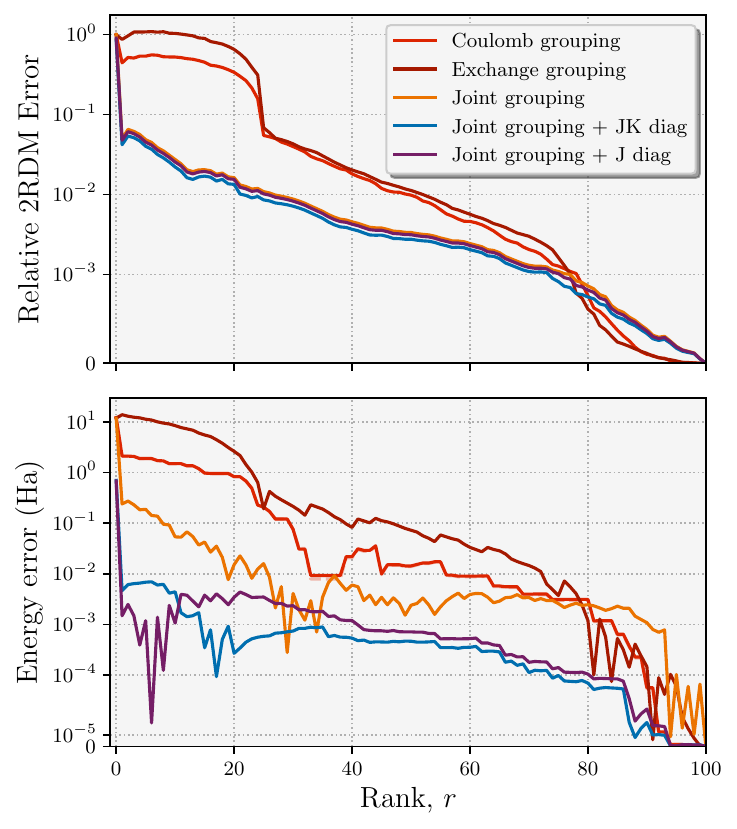}
  \caption{Low-rank decomposition of FCI two-body reduced density matrix of H$_{10}$ in a minimal STO-6G basis at $1.5$a$_0$ interatomic separation, using different groupings of orbital indices, with and without diagonal corrections. Diagonal corrections augmenting the joint decomposition involving only Coulomb diagonals ($J$) as well as all three Coulomb and exchange diagonals ($JK$) are shown. Upper plot characterizes the relative mean absolute error over all 2RDM elements, while the lower plot characterizes the absolute two-electron energy error from the reconstructed 2RDM.}
  \label{fig:fci-decomp}
\end{figure}

We now move beyond formally low-rank 2RDMs and consider the construction of effective, compact, and systematically improvable descriptors for general correlated 2RDMs, for which the rank required for an exact representation is expected to grow with system size. In Fig.~\ref{fig:fci-decomp}, we therefore examine low-rank truncations of full configuration interaction (FCI) ground-state 2RDMs for the linear H$_{10}$ chain at 1.5~a$_0$, close to its equilibrium geometry, obtained with the \textsc{PySCF} package~\cite{pyscf2018,pyscf2020}. We report both the relative norm error in the reconstructed 2RDM and the resulting error in the two-electron contribution to the energy as functions of the retained rank. As expected, exact reconstruction of the FCI 2RDM requires the full rank. Nevertheless, the joint decomposition achieves significantly smaller errors at fixed rank than the single-channel decompositions. In particular, an energy accuracy of 10~mHa is reached at approximately 20\% of the full rank in the joint decomposition, compared to roughly 35\% and 70\% in the Coulomb and exchange groupings, respectively. Near full rank this trend is partially reversed, with the single-channel decompositions converging slightly more rapidly in the final approach to accuracies below $10^{-4}$~Ha, but this is the only regime in which they are favored.

However, not all elements of a 2RDM are equally important for the faithful reconstruction of many expectation values of interest. In particular, relatively small errors in certain structured subsets of tensor elements can have a disproportionate effect on physical properties, even when the global norm error remains modest. The simplest example is the normalization condition
\begin{equation}
\sum_{ij} \Gamma_{iijj} = N(N-1),
\end{equation}
which fixes the particle number and directly affects the interaction energy. This quantity is not generally preserved by the low-rank truncations described above. More generally, diagonal slices of the 2RDM also play an important role in characterizing two-point spin and charge correlators~\cite{https://doi.org/10.1002/qua.24101}.

To reduce these errors in our compressed form, we can augment the truncated low-rank representation by an explicit post-truncation correction to diagonal elements of the 2RDM. Consistent with the three inequivalent interaction channels introduced in Sec.~\ref{sec:singlechanneldecomps}, this gives rise to three distinct diagonal slices, namely $\Gamma_{iijj}$, $\Gamma_{ijij}$, and $\Gamma_{ijji}$. The first corresponds to the diagonal in the Coulomb-like grouping, while the latter two arise from exchange-like permutations of the orbital indices. Although only $\Gamma_{iijj}$ directly enforces the particle number sum rule, correcting all three diagonals improves the fidelity of many observables derived from the compressed 2RDM, including local spin-sensitive quantities.

Unlike the spectral decompositions themselves, these diagonal corrections are not invariant under orbital rotations. Their effectiveness therefore depends on the basis in which the correction is imposed. We choose to define these corrections in a local orthogonalized atomic-orbital basis, motivated by the expectation that in large systems the most important residual errors are associated with local two-body fluctuations, which remain non-sparse in the asymptotically large system size limit. A Löwdin-orthogonalized atomic-orbital basis provides a natural compromise between locality and orthogonality, and is widely used in local correlation analyses~\cite{doi:10.1021/acs.accounts.4c00085}. We therefore define the symmetrically orthogonalized atomic orbitals (SAOs) as
\begin{equation}
\label{eq:sao}
\chi_i
=
(s^{-1/2})_{wi}\,\phi_w,
\end{equation}
where $s_{wx}=\langle \phi_w | \phi_x \rangle$ is the overlap matrix of an underlying atomic-orbital basis. Imposing exact diagonal corrections in this representation is designed to preferentially improve local charge and spin correlators, as well as correlation effects dominated by local fluctuations, e.g. in stretched bonds, transition states or the Mott character of high-$U$ Hubbard models, where these corrections will ensure the accuracy of local double occupancies. This basis choice is also consistent with the eigenvector continuation applications discussed in Sec.~\ref{sec:NucEnsem}.

The SAO diagonal corrections are stored as three matrices,
\begin{align}
    D^1_{ij} &= \Gamma_{iijj} - \Gamma_{iijj}^{\mathrm{lowrank}}, \label{eq:diagJ}\\
    D^2_{ij} &= \Gamma_{ijij} - \Gamma_{ijij}^{\mathrm{lowrank}}, \label{eq:diagK1}\\
    D^3_{ij} &= \Gamma_{ijji} - \Gamma_{ijji}^{\mathrm{lowrank}}, \label{eq:diagK2}
\end{align}
where $\Gamma^{\mathrm{lowrank}}$ denotes the truncated spectral representation of the 2RDM. Including these matrices in the compressed form ensures that all three diagonal slices are reproduced exactly in the SAO basis. Each SAO correction matrix has the same memory cost as a single low-rank vector, yet yields a substantial improvement in energetics. 

The resulting approximation can be interpreted as a constrained low-rank reconstruction of the 2RDM. The low-rank component accounts for the dominant long-range and nonlocal correlation structure, while a small additive correction enforces selected diagonal marginals exactly. This viewpoint echos recent work on low-rank matrix completion and noise filtering strategies for the 2RDM, where incomplete or noisy data are regularized by imposing a compact low-rank structure together with exact consistency conditions on selected entries or channels~\cite{doi:10.1021/acs.jpclett.6c00296,Chan23-RDM-completion,Candes2009}. Here, the additional constraints are motivated directly by the physics of the observables rather than by missing data.

The effect of applying only the Coulomb-like diagonal correction $D^1_{ij}$ ($J$) and of applying all three diagonal corrections ($JK$) is shown in Fig.~\ref{fig:fci-decomp}. With these corrections included, an energy accuracy of 1~mHa is reached with only a single retained vector, representing more than an order-of-magnitude reduction in the rank required relative to the uncorrected decomposition. Because these diagonal corrections account for only a vanishing fraction of the full 2RDM elements, their effect on the global norm error remains modest; their principal value lies instead in the improved accuracy of physically relevant observables.
%It is interesting to note that J diagonal corrections by themselves result in a more compressed representation for $10^{-4}$~Ha accuracy compared to including all diagonal corrections. This trend reverses, however, where all diagonal case slightly overperforms J diagonals after it leaves its initial trough.

\subsection{Amplitude relaxation} \label{sec:relax}

We can further improve the low-rank representation of the 2RDM by relaxing the linear coefficients of the joint decomposition expansion of Eq.~\ref{eq:joint_decomp}, $\varepsilon^{(\alpha)}$, while keeping the vectors $v^{(\alpha)}$ fixed. This aims to ensure that the expansion best approximates the original 2RDM, rather than the auxiliary variable, $Q_{ijkl}$, in a least-squares sense. The $\varepsilon^{(\alpha)}$ enter the ansatz linearly, which allows us to define an objective function of the 2RDM frobenius norm error for the low-rank decomposition we wish to minimize, as
\begin{equation}
      \min_{\varepsilon}
  \left\|
    \Gamma_{x} -
    \sum_{\alpha=1}^r \epsilon^{(\alpha)} B_x^\alpha 
  \right\|_F^2 , \label{eq:costfn}
\end{equation}
where $B_{ijkl}^\alpha$ are the spin-free wedge-product basis functions of the expansion,
\begin{equation}
    B_{ijkl}^\alpha = v_{ij}^{(\alpha)} v_{kl}^{(\alpha)} - \frac{1}{2} v_{il}^{(\alpha)} v_{kj}^{(\alpha)} ,
\end{equation}
and $x$ in Eq.~\ref{eq:costfn} denotes the compound indices of the four-index elements of the 2RDM which the fit aims to optimize over. We can solve Eq.~\ref{eq:costfn} in closed form in a least squares sense, by forming the (generally rank-deficient) Gram matrix
\begin{equation}
    G_{\alpha \beta} = \sum_x B_x^\alpha B_x^\beta
\end{equation}
and solving the linear equation for the vector of relaxed amplitudes, $\varepsilon^{(\beta)}$, as
\begin{equation}
    G_{\alpha \beta} \varepsilon^{(\beta)} = b_\alpha ,
\end{equation}
where the right hand side is given by
\begin{equation}
    b_\alpha = \sum_x B_x^\alpha \Gamma_x .
\end{equation}
If combining the amplitude relaxation with a diagonal correction, then the $x$ indices will exclude these diagonal contributions, and the final diagonal correction can be found after the relaxation via Eqs.~\ref{eq:diagJ}-\ref{eq:diagK2}, using the relaxed coefficients in the low-rank approximation. The approach could also be used to optimize the linear coefficients for contracted quantities, including the 1RDM or other variables in a similar fashion, if particular importance was placed to certain expectation values rather than assigning equal importance to all 2RDM elements. This relaxation has more impact when used with the diagonal corrections, as the coefficients can be relaxed to account for the fact that they are not trying to fit these diagonal entries of the 2RDM. Benchmarking the improvement of the low-rank expansion from the relaxation of these linear amplitudes is shown in Appendix~\ref{app:relax}. However, the approach ultimately had a relatively small impact on the quality of the low-rank approximations, and so has not been used unless mentioned otherwise in the remaining results of this paper.

\subsection{Transition RDMs}
\label{sec:tRDM}

The extension of the low-rank decomposition to transition two-body reduced density matrices follows a similar physical motivation. The joint channel low-rank form for single determinants as given in Eq.~\ref{eq:hf-2rdm} still exhibits the wedge-product structure, suggesting that the joint decomposition of Eq.~\ref{eq:joint_decomp} can remain effective, absorbing the ${}^{ab}S^{-1}$ factors of Eq.~\ref{eq:hf-2rdm} into the $\varepsilon^{(\alpha)}$ values of the decomposition. Care needs to be taken in cases where ${}^{ab}S \rightarrow 0$, but this can be dealt with effectively as shown in Appendix~\ref{app:trdm}, where the 2tRDM between general determinants is derived even for cases where ${}^{ab}S \rightarrow 0$. Furthermore, we derive the 2tRDMs from the generalized Slater--Condon expressions, providing a robust procedure for their implementation so that quantities remain finite, and detail the formal rank for a joint decomposition form of these transition quantities for different number of orbital excitations in the two states. 
Numerical tests analogous to those of Fig.~\ref{fig:fci-decomp} indeed show that the joint decomposition outperforms individual channel decompositions, and that diagonal corrections again improve the corresponding transition-energy error metrics. The quantities ${}^{ab}\Gamma_{ijkl}$ (and the analogous ${}^{ab}Q_{ijkl}$ object of Eq.~\ref{eq:coulombvec_jointdecomp}) remain Hermitian for this transition case.

%% file: 4-results2.tex
\section{Scaling and low-rank contraction}
\label{sec:scaling_inference}

\subsection{System size scaling}
\label{sec:alkanes}

\begin{figure*}[htb]
\centering
\includegraphics[width=0.99\linewidth]{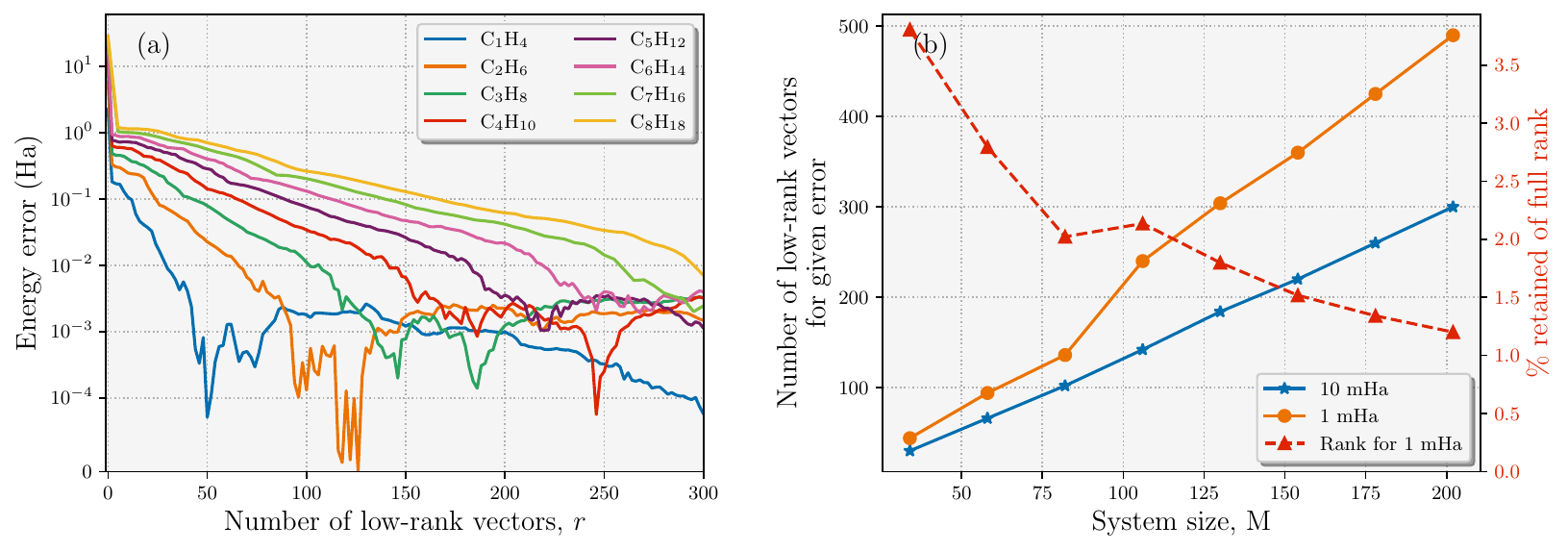}
\caption{System size scaling of the low-rank compression. (a) Total energy error introduced by the joint decomposition of CCSD 2RDMs of alkane chains in a cc-pVDZ basis with $J$ diagonal corrections. (b) Number of vectors required for a desired total energy error of the decomposition as the number of orbitals in the system increases. %Two-electron integrals were approximated via density-fitting with the cc-pVDZ-RI auxillary basis. 
}
%, (c) inference time scaling (energy+force) using low-rank decompositions of xxx.}
\label{fig:scaling}
\end{figure*}

We next consider how the rank required for an accurate compressed 2RDM representation scales with system size. The full-rank decomposition scales quadratically with the number of orbitals, \(M\). By contrast, for general correlated states we expect the rank required to achieve a fixed target accuracy to grow much more slowly, and in particular to display an approximately linear scaling in systems, consistent with the extensivity of the irreducible two-body correlations of the cumulant~\cite{10.1063/1.478189}. To test this, we study truncated joint decompositions with Coulomb-channel diagonal corrections in cases where no exact low-rank form is expected. We consider the ground-state 2RDMs of alkane chains of increasing length, \(\mathrm{C}_{n}\mathrm{H}_{2n+2}\), in the cc-pVDZ basis, using CCSD to access larger systems~\cite{pyscf2018,pyscf2020}. The molecular geometries were optimized with the \textsc{geomeTRIC} package~\cite{10.1063/1.4952956}.

Figure~\ref{fig:scaling}(a) shows the error in the energy recovered from the truncated joint decomposition for alkane chains up to octane, corresponding to systems with up to 202 orbitals. To quantify the scaling of the compression more directly, Fig.~\ref{fig:scaling}(b) reports the number of retained vectors required to reach total energy target accuracies of 10~mHa and 1~mHa. To reduce sensitivity to small numerical fluctuations, we define the rank required to achieve a given threshold as the smallest retained rank for which the target error is satisfied for at least two consecutive increases in truncated rank.

As anticipated, the results clearly show approximately linear growth in the number of retained vectors required to reach a fixed energy accuracy, in contrast to the quadratic growth of the full rank. The slope of this linear trend increases as a more stringent energy target is imposed. This indicates that the cost of maintaining a fixed absolute accuracy grows extensively with system size, rather than with the full quadratic dimension of the 2RDM pair space. Equivalently, the number of vectors required per electron, or per unit volume, remains approximately constant across the alkane series, leading to a quadratically scaling 2RDM memory requirement with system size for a fixed relative energy error. A similar extensive scaling was previously observed in low-rank parametric 2RDM approaches that coupled multiple decomposition channels~\cite{Mazziotti12-lowrank}. Despite this scaling, much larger ranks were found to be required to obtain substantially sub-mHa accuracy in Fig.~\ref{fig:scaling}, with a convergence profile which appears to plateau for a while at this accuracy before reducing further.

%\ghb{Shall we say something about the fact that it seems 'hard' to get the total energy error below 1mHa, regardless of system size? Its mentioned later, but perhaps worth a note here.}

The prefactor of this scaling is also important in practice. The right-hand axis of Fig.~\ref{fig:scaling}(b) therefore shows the retained rank as a percentage of the full rank, providing a direct measure of the achieved compression. Because the full rank grows quadratically while the retained rank grows only linearly, the retained fraction decreases approximately inversely with system size. For the largest system studied here, octane (\(\mathrm{C}_8\mathrm{H}_{18}\)) with 202 orbitals, only 490 low-rank vectors are required to achieve 1~mHa accuracy in the inferred energy, compared to 40\,804 vectors at full rank, corresponding to a compression of 98.7\%. This demonstrates that the proposed representation becomes increasingly efficient as the system size grows. This behavior is consistent with the extensive nature of correlation, for which the irreducible two-body structure is expected to grow approximately linearly with system size. It however does not rely on locality {\em per se}, and the rank-scaling observed from the joint decomposition is independent of whether the RDMs are expressed in a local basis or not. However, in a local basis, further emergent sparsity is expected within the individual vectors, $v_{ij}^{(\alpha)}$, leading asymptotically to an $\mathcal{O}(M)$ scaling in their information content, potentially motivating future approaches based on local truncations to reduce the memory scaling further. 

%The number of vectors required to reach 1~mHa accuracy also scales linearly, with a piecewise discontinuity between $\mathrm{C}_3\mathrm{H}_8$ and $\mathrm{C}_4\mathrm{H}_10$, with 82 and 106 orbitals respectively. This can be understood from the early double trough nature of the energy error for these decompositions in Fig.~\ref{fig:scaling}(a). The initial linear relationship corresponds to the first trough reaching below the accuracy of 1~mHa, whereas only the second trough reaches that, $\mathrm{C}_4\mathrm{H}_{10}$ onwards. A marginally steeper linear scaling is observed for the second trough that is captured after $\sim$100 orbitals, indicating a linearly increasing separation of the two troughs in the energy error.

\subsection{Computation with low-rank vectors}
\label{sec:inference}

%\ghb{Are we going to insert back the $\varepsilon^{(\alpha)}$ eigenvalues explicitly into the form, derived from Eq.~\ref{eq:coulombvec_jointdecomp}, rather than \ref{eq:joint_decomp}?} \ka{Done, also changed Eq. \ref{eq:joint_decomp}}

While the memory reduction afforded by the compression is valuable, the representation is only practically useful if two-body observables can also be evaluated directly from the compressed form, without reconstructing the full 2(t)RDM. In this section, we show how this can be done for the energy and nuclear gradients within the joint decomposition, with further details for the diagonal correction given in Appendix~\ref{app:inference}. The idea is to exploit standard efficient atomic-orbital (AO) Coulomb and exchange matrix builds, as implemented in many Hartree--Fock and hybrid-DFT routines, after transforming each retained low-rank vector to the AO basis,
\begin{equation}
v_{wx}^{(\alpha)} = Z_{wi} Z_{xj} v_{ij}^{(\alpha)},
\end{equation}
where \(w,x,y,z\) denote AO indices and \(Z\) is the orbital-to-AO transformation matrix.

%\subsubsection*{Energy}
Using the joint decomposition of Eq.~\ref{eq:joint_decomp}, the two-electron contribution to the energy can be evaluated as
\begin{equation}
E_{ee}
=
\frac{1}{2}
\sum_{\alpha} \varepsilon^{(\alpha)}
\left[
v_{wx}^{(\alpha)} v_{yz}^{(\alpha)} (wx|yz)
-
\frac{1}{2}
v_{wz}^{(\alpha)} v_{yx}^{(\alpha)} (wx|yz)
\right],
\end{equation}
with Einstein convention assumed.
This can be evaluated with AO Coulomb and exchange builds as
\begin{equation}
E_{ee}
=
\frac{1}{2}
\sum_{\alpha} \varepsilon^{(\alpha)}
\left[
v_{wx}^{(\alpha)} J_{wx}^{(\alpha)}
-
\frac{1}{2}
v_{wx}^{(\alpha)} K_{wx}^{(\alpha)}
\right],
\label{eq:jointED_jkbuild}
\end{equation}
where these Coulomb and exchange matrix builds are defined by contraction of the AO two-electron integrals with each retained low-rank vector,
\begin{equation}
J_{wx}^{(\alpha)} = (wx|yz) v_{yz}^{(\alpha)},
\qquad
K_{wx}^{(\alpha)} = (wz|yx) v_{yz}^{(\alpha)} .
\label{eq:jkbuilds}
\end{equation}
In practice, this has the same computational structure as the conventional Coulomb and exchange-matrix builds of Hartree--Fock or hybrid-DFT, but with the (generally non-Hermitian) low-rank vector \(v^{(\alpha)}\) replacing the usual one-body density matrix. This is not a fundamental obstacle, however, since the underlying AO integral contractions remain unchanged and most implementations can be generalized straightforwardly to accept nonsymmetric input matrices.

While a naive evaluation of Eqs.~\ref{eq:jkbuilds} scale as \(\mathcal{O}(M^4)\) with the AO basis size \(M\), in practice these $J/K-$builds are among the most highly optimized kernels in electronic-structure theory, and a range of established approaches -- including density fitting / resolution-of-the-identity, pseudospectral methods, tensor hypercontraction, and multipole-based techniques -- can reduce the cost and scaling substantially~\cite{10.1063/1.2956507,doi:10.1021/acs.jctc.9b00228,doi:10.1021/acs.jctc.8b00358,10.1063/1.473833,10.1063/1.2244565}. In the present work, we employ density fitting with a cc-pVDZ-RI auxiliary basis to reduce prefactors and obtain \(\mathcal{O}(M^3)\) scaling for the Coulomb builds. The overall cost of evaluating the two-electron energy from the compressed representation is therefore the cost of a single \(J/K\)-type build multiplied by the number of retained low-rank vectors.

%\ghb{I think we should add something about the diagonal corrections here - at least pointing to the fact that they can be included simply, and to see the appendix B for further details.}

%\subsubsection*{Nuclear gradients}

The same AO Coulomb- and exchange-build machinery can also be used to evaluate nuclear gradients directly from the compressed representation. Assuming that the underlying 2RDM is obtained from a stationary electronic state, so that explicit response of the wavefunction parameters does not contribute at first order, the two-electron part of the nuclear gradient is
\begin{equation}
\nabla_{\mathbf R} E_{ee}
=
\frac{1}{2}\,
\Gamma_{ijkl}\,
\nabla_{\mathbf R}(ij|kl),
\label{eq:gradE}
\end{equation}
where the nuclear derivative of the molecular-orbital electron-repulsion integral contains both derivatives of the underlying AO integrals and derivatives of the AO-to-orbital transformation~\cite{https://doi.org/10.1002/wcms.1171,doi:https://doi.org/10.1002/9780470749593.hrs006}.

%Here, we assumed a static 2RDM which is the case in the eigenvector continuation scheme that will be discussed in Sec.~\ref{sec:eigcon}. 
%In this scheme, 2RDM is fixed and the two-body correlations it captures are transferred by the changing single particle basis that is captured by the $\nabla_\mathbf{R}Z_{wi}$ contributions.
%+\frac{1}{2} (wx|yz) \left(\nabla_{\mathbf R} \Gamma_{wxyz}\right)

The contribution arising from AO integral derivatives may be written as
\begin{equation}
\nabla_{\mathbf R} E^{(\mathrm{eri})}
= \frac{1}{2}
\left(
\frac{\partial w}{\partial \mathbf R} x \Big| yz
\right)
\Big[\Gamma_{wxyz}+\Gamma_{xwzy} + \Gamma_{yzwx} + \Gamma_{zyxw}\Big],
\label{eq:eri_grad}
\end{equation}
where the four terms account for differentiation of each AO slot in the electron-repulsion integral through the usual permutation symmetry. Substituting the joint decomposition of Eq.~\ref{eq:joint_decomp} into Eq.~\ref{eq:eri_grad} and regrouping equivalent Coulomb- and exchange-like contributions yields a sum of AO derivative contractions that has the same structure as a $J/K-$build derivative for each retained vector. Denoting the resulting derivative build for an input matrix \(D\) by \(\nabla_{\mathbf R}F(D)\), the ERI-derivative contribution can be written compactly as
\begin{align}
\nabla_{\mathbf R} E^{(\mathrm{eri})}
=
\sum_{\alpha} \varepsilon^{(\alpha)}
\Big[
&v_{wx}^{(\alpha)}\, \nabla_{\mathbf R}F_{wx}(v^{(\alpha)T}) \nonumber
\\ 
+ \,
&v_{xw}^{(\alpha)}\, \nabla_{\mathbf R}F_{wx}(v^{(\alpha)})
\Big],
\label{eq:eri_grad_joint}
\end{align}
where we denote the $J/K-$derivative builds of a one-body object, $D$, as
\begin{equation}
\nabla_{\mathbf R} F_{wx}(D)
=
\left(
\frac{\partial w}{\partial \mathbf R} x \big| yz
\right)
D_{zy}
-
\frac{1}{2}
\left(
\frac{\partial w}{\partial \mathbf R} z \big| yx
\right)
D_{zy},
\end{equation}
once again allowing us to reuse performant mean-field subroutines~\cite{10.1063/1.2906127,10.1063/5.0131683}. %\ghb{This above equation looks wrong dimensionally and index-wise. Can you check this carefully (or we can remove...?).} \ka{See the change - I took the indexing straight from pyscf function comment, also happy to remove it if we assume reader can extrapolate from J/K builds to gradient J/K builds.}

%Substituting the joint decomposition in Eq.~\ref{eq:joint_decomp} into this expression gives:
%\begin{align}
%\nabla_{\mathbf R} E^{(\mathrm{eri})}
%&
%=
%\nonumber \\
%\sum_{\alpha}
%\Big[
%& v_{wx}^{(\alpha)}\left(\frac{\partial w}{\partial \mathbf R} x \big| yz\right)v_{yz}^{(\alpha)} 
%-\frac{1}{2}
%v_{wz}^{(\alpha)}\left(\frac{\partial w}{\partial \mathbf R} x \big| yz\right)v_{yx}^{(\alpha)}\nonumber\\
%+
%&v_{xw}^{(\alpha)}\left(\frac{\partial w}{\partial \mathbf R}x\big|yz\right)v_{zy}^{(\alpha)}
%-\frac{1}{2}
%v_{zw}^{(\alpha)}\left(\frac{\partial w}{\partial \mathbf R} x \big|yz\right)v_{xy}^{(\alpha)}\Big].
%\end{align}
%where contributions from $wxyz$ ($xwzy$) and $yzwx$ ($zyxw$) permutations are grouped, as they are equivalent to each other due to the permutation symmetry of the representation.

In addition, there are contributions from differentiation of the orbital transformation itself. These are analogous to the usual Pulay terms associated with atom-centered basis functions. Using the same two-body Fock-like intermediates defined for the energy evaluation, they can be written as
\begin{equation}
\nabla_\mathbf{R} E^\mathrm{(orb)}
=
\sum_\alpha \varepsilon^{(\alpha)}
\left(\nabla_\mathbf{R}Z_{wi}\right) Z_{xj}
\left[
v_{ij}^{(\alpha)} F_{wx}^{(\alpha)}
+
v_{ji}^{(\alpha)} F_{xw}^{(\alpha)}
\right],
\label{eq:pulay_joint}
\end{equation}
where the full derivation is given in Appendix~\ref{app:inference}.
The cost of evaluating the orbital-response term is cheap, since it allows reuse of the same $J/K$ intermediates evaluated for the energy. The derivative integral builds are somewhat more expensive, but inherit the same asymptotic scaling as the corresponding energy builds when density fitting is employed and the response of the auxiliary basis is neglected. The strategy of reusing mean-field AO-driven functionality extends analogously to all two-body observables, including non-adiabatic couplings from the low-rank decomposition of transition two-body density matrices, as implemented for Sec.~\ref{sec:namd}.

%% file: 5-results-evcont.tex
\section{Across atomic geometries} \label{sec:NucEnsem} %\ghb{I think we probably want to clarify this title a little - something like 'Towards application in eigenvector continuation, with a subtitle of 'RDM decomposition across geometries'}\ka{I think it's fine to leave the section title like this, or 'across nuclear/molecular geometries'. I restructured the section so subspace Hamiltonian is introduced first, and the 'transition energy' is motivated better through this form}

\subsection{RDM compression across geometries} \label{sec:compresstraining}

We next assess the performance of the compression scheme across varying molecular conformations and correlation regimes, in the context of an \emph{ab initio} wavefunction interpolation framework recently introduced by the authors~\cite{Rath25-evcont,Atalar24-eigconNAMD-faraday}. This interpolation is based on eigenvector continuation, as developed in nuclear and condensed-matter physics~\cite{PhysRevLett.121.032501,RevModPhys.96.031002}, and allows for a set of training wavefunctions computed at representative geometries to be used to span a low-dimensional electronic subspace across the continuous manifold of nuclear configurations. At each new geometry, the electronic state is then approximated by solving the Schr\"odinger equation variationally within this subspace, yielding an optimal linear combination of the training state vectors in a particular local representation.

Within this framework, the one- and two-body transition reduced density matrices between training states \(a\) and \(b\) are required to project the Hamiltonian into the training-state subspace and compute the individual matrix elements. The resulting subspace Hamiltonian at geometry \(\mathbf{R}\) is
\begin{align}
    \mathcal{H}_{ab}(\mathbf{R}) &= \sum_{ij} {}^{ab}\gamma_{ij}\, h_{ij}(\mathbf{R}) \nonumber \\
    &\quad + \frac{1}{2} \sum_{ijkl} {}^{ab}\Gamma_{ijkl}\, (ij|kl)(\mathbf{R})
    + E_{\mathrm{nuc}}(\mathbf{R}),
    \label{eq:subspaceh}
\end{align}
so that the training wavefunctions themselves need not be retained explicitly once the corresponding (transition) RDMs have been constructed. The essential approximation is that the many-body amplitudes of the training states define a transferable diabatic-like basis in which the wavefunction at nearby geometries can be accurately represented. In the present work, both the (transition) RDMs and the Hamiltonian are represented in the SAO basis of Eq.~\ref{eq:sao}, which improves computability and transferability across related molecular geometries~\cite{Rath25-evcont,Atalar24-eigconNAMD-faraday,mejutozaera2023-evcont}.

Although the inference step at a test geometry formally requires \(\mathcal{O}(N_{\mathrm{train}}^2 M^4)\) compute to evaluate Eq.~\ref{eq:subspaceh} over all pairs of training states, one of the dominant practical bottlenecks is the storage of the \(\mathcal{O}(N_{\mathrm{train}}^2)\) transition 2RDMs themselves, each of $\mathcal{O}(M^4)$ size. For realistic basis sizes and training-set cardinalities, this quartic memory cost in \(M\) rapidly becomes prohibitive. The compression scheme developed here reduces this storage requirement to \(\mathcal{O}(r\,N_{\mathrm{train}}^2 M^2)\), where \(r\) is the retained rank of the compressed representation, thereby substantially extending the range of systems and training manifolds accessible to the eigenvector-continuation workflow.

We first consider the efficiency of the low-rank compression across various nuclear geometries in addition to the equilibrium geometries considered previously, which will subsequently be used as training states for the interpolation in Sec.~\ref{sec:inference}. The density matrices used for this analysis are computed from matrix product states (MPS) using DMRG as implemented in \emph{block2}\cite{Block2} package, for the H$_{28}$ molecule in minimal basis. 44 states across 22 linear geometries (including the ground and the first singlet excited many-body state on each) are optimized, which were selected from the nuclear phase space explored from a photo-excited fewest switches surface hopping non-adiabatic molecular dynamics trajectory starting from a symmetric initial configuration, as described in our previous work~\cite{Atalar24-eigconNAMD-faraday}. The MPS of these different geometries were optimized directly in the SAO basis, systematically increasing the bond dimension under further increases yielded negligible energy improvements in either the ground or $S_1$ excited states, and subsequently used to generate the 2(t)RDMs between all these states. In this way, we can be confident of the near-FCI accuracy of the states allowed for by DMRG for this one-dimensional molecular topology with a local basis, beyond the capabilities of FCI, while not expecting a significant affect of this restricted topology on the compactness of the representation or conclusions we can reach.

There are 990 unique (transition) 2RDMs between the 44 many-body states and we study the rank of the expansion required to achieve a target absolute total energy error from the contraction of these (t)RDMs. For the non-transition RDMs, this error can be calculated simply from the difference in Eq.~\ref{eq:energyfunc} between the compressed and uncompressed RDMs, while for the transition RDMs we consider the `transition' energy defined as the contraction of the tRDM with the two-electron integrals of the bra state geometry - analogous to the contraction in the subspace Hamiltonian in Eq.~\ref{eq:subspaceh}. We note that the one-body energy contributions are obtained from the exact 1(t)RDMs, rather than contracting the compressed representation via Eq.~\ref{eq:1trdm}. All the (t)RDMs in the rest of this work are compressed using joint decomposition with $J$ diagonal corrections applied.

\begin{figure}[tbh]
    \centering
    \includegraphics[width=0.97\linewidth]{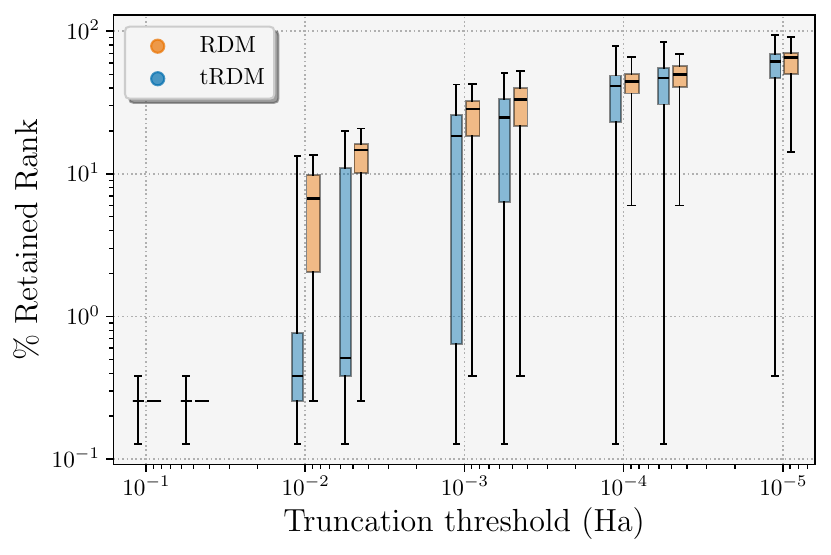}
    \caption{Compressed rank of the low-rank non-transition and transition 2-body reduced density matrices of $H_{28}$ chains in minimal basis compared to their full rank (exact) form, for different maximum introduced total energy error. The distribution is shown for the 44 $S_0$ and $S_1$ many-body states across 22 geometries, and the resulting compressed 990 (transition) 2RDMs at each energy truncation threshold. The box is defined by the 25th and 75th percentile of the distribution and the median is shown by a line in the middle, while the whiskers highlight the minimum and maximum values of the distribution.}
    \label{fig:h28-distributions}
\end{figure}

By initially computing the two-electron energy of the full 2(t)RDM, the rank can then be systematically increased until a desired target accuracy in the total energy error is achieved.
The fraction of the retained rank in the compressed 2(t)RDMs compared to the full rank for both non-transition and transition RDMs over different energy truncation thresholds is shown in Fig.~\ref{fig:h28-distributions}, given as a distribution across a diverse range of geometries and correlation regimes explored in a molecular dynamics trajectory.
Firstly, one can notice that the median rank is consistently lower for the transition RDMs despite the maximum rank being similar in both transition and non-transition RDMs. As discussed earlier, joint decomposition performs very similarly to non-transition RDMs when the overlap between states are close to one. This is why the maximum rank shown by the top whiskers in Fig.~\ref{fig:h28-distributions} is similar for both non-transition and transition RDMs. For low overlap transition RDMs, however, the two-body correlations are more sparse allowing them to be captured with more compact representations. This can bring the tRDMs rank much lower as seen for the energy truncation thresholds around $10^{-2}$~Ha and $10^{-3}$~Ha in Fig.~\ref{fig:h28-distributions}.

It is also noticeable that there is a significant spread in the compression achieved across geometries. Both the min-max separation and interquartile ranges (IQR) are large for certain thresholds, sometimes spanning orders of magnitude in compressed rank. This variation is larger for transition RDMs due to overlap between states varying from near-one to strictly zero (between ground and excited states) at different geometries, as seen by the IQR of the tRDM distributions being a factor of 2-3 larger than those of the non-transition RDMs. The IQR of the non-transition RDM ranks is roughly 10\% across most truncation thresholds, showing a consistent compression across geometries and an overall median compression of $\sim$70\% to achieve chemical accuracy, even for this relatively small system size.

\subsection{Effect of RDM compression on interpolation}
\label{sec:eigcon}

A central aim of this work is to apply the low-rank compression at the \emph{training} stage, so that the full set of training 2(t)RDMs can be stored and reused more efficiently during subsequent interpolation at unseen geometries. Because the rank required for a given truncation varies substantially across geometries and transition channels, it is important to establish that the metric used to choose this rank at the training stage also provides reliable control over the final interpolation error. In other words, the local truncation criterion used in decomposing the training 2(t)RDMs must transfer robustly to the accuracy of the inferred energies at test geometries.

%One of the aims of this work was to apply the low-rank compression at the `training' stage, systematically decomposing the training 2(t)RDMs to enable a more efficient subsequent interpolation at unseen test geometries. However, since the rank of the decomposition varies significantly between geometries, we need to ensure that the metrics used to decide on this rank (e.g. in Fig.~\ref{fig:h28-distributions}), transfer over to the accuracy of the interpolation, so that robust and controllable truncations of the 2(t)RDMs can be performed across to the resulting inference stage.

The subspace Hamiltonian of Eq.~\ref{eq:subspaceh} can be assembled efficiently from the low-rank vectors using the AO-based JK-build strategy described in Sec.~\ref{sec:inference}. The interpolated energies at geometry \(\mathbf{R}\) are then obtained from the generalized eigenvalue problem
\begin{equation}
    \left(S_{ab}^{-\frac{1}{2}} \, \mathcal{H}_{ab} \, S_{ab}^{-\frac{1}{2}}\right) \mathbf{C}_\mu = E_\mu \mathbf{C}_\mu,
\label{eq:ortho_subspaceh}
\end{equation}
%\textcolor{red}{(HB: I think some more clarity is needed here on the definition of $S_{ab}^{-1/2}$. Is this the overlap of many-body basis states, or transformation in the orthogonalized AO basis? Also, this is similar notation that used for the many-body overlap ${}^{ab}S$ in the appendix, which could be confusing if it is in fact the AO overlap being considered here. Same goes for Eq.~\eqref{eq:ortho_rdm} below)}
where \(\mathbf{C}_\mu\) gives the contribution of each training state to the variationally optimal interpolated state, and $S_{ab}$ denotes the overlap between the many-body training states labelled by $a$ and $b$. In practice, for numerical stability, the orthogonalization within the training manifold is applied to the RDMs before compression,
\begin{equation}
    \tilde\Gamma_{ijkl}^{ab} = S_{ab}^{-\frac{1}{2}} \, \Gamma_{ijkl}^{ab} \, S_{ab}^{-\frac{1}{2}},
\label{eq:ortho_rdm}
\end{equation}
and it is these orthogonalized objects, \(\tilde\Gamma_{ijkl}^{ab}\), that are subsequently decomposed.

\begin{figure}[htb]
\centering
  \includegraphics[width=\linewidth]{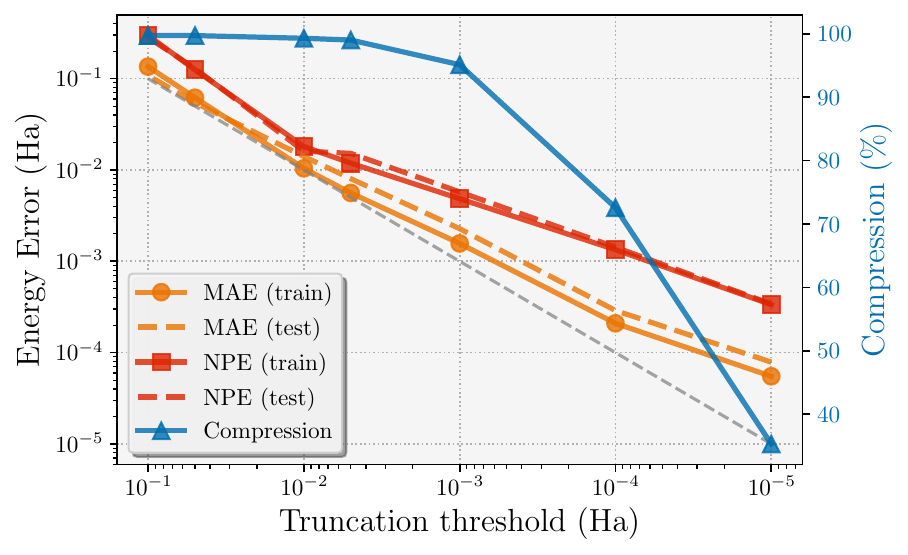}
  \caption{Effect of low-rank compression of the $S_0$ and $S_1$ interpolated energies over the `training' and `test' geometries sampled along a nonadiabatic molecular dynamics trajectory. Mean absolute and non-parallelity energy errors are shown for different truncation thesholds with respect to full continuation, with 44 many-body MPS training states of H$_{28}$ in a minimal basis. Right $y$-axis shows the compression rate of the truncation relative the full 2-body RDM. Dashed line shows $x$=$y$ for the correspondence between truncation threshold and energy error.}
  \label{fig:h28_hamthr}
\end{figure}

Using the 44 \(S_0\) and \(S_1\) states of Sec.~\ref{sec:compresstraining} as a fixed training basis, Fig.~\ref{fig:h28_hamthr} examines how truncation of the corresponding 2(t)RDMs affects the interpolated energies at both training and test geometries. The test geometries are taken from a short nonadiabatic molecular dynamics trajectory generated with fewest-switches surface hopping (FSSH), initialized in the \(S_1\) state near the ground-state equilibrium geometry, generating 80 test states allow the trajectory to assess their accuracy; further details are given in Sec.~\ref{sec:namd}. For each truncation threshold, all training RDMs are compressed to satisfy a chosen target error in the two-body energy metric at the point of decomposition. %Since the orthogonalized RDMs in Eq.~\ref{eq:ortho_rdm} now correspond to linear combinations of states from different geometries, we arbitrarily choose to contract these RDMs by the two-electron integrals from different training geometries in their original ordering to obtain the two-body energy metric used for truncation.  \ghb{How is the Hamiltonian / error metric chosen for the orthogonalized states - needs a bit more explanation} \ka{The integrals are selected in the original order of non-orthogonal 2RDMs. Probably should've chosen the ones from the geometry with highest contribution but I don't think it would've made a big difference. Added an explanation for it.}. 
The resulting interpolation errors are then separately evaluated over the original training geometries and over the unseen test geometries, for both the \(S_0\) and \(S_1\) potential energy surfaces. We report both the mean absolute error (MAE) and the non-parallelity error (NPE) in the total energy, where the NPE is defined as
\begin{equation}
\mathrm{Err}_{\mathrm{NPE}}=\max_{\mathbf{R}}(E_{\mathrm{full}} - E_{\mathrm{lowrank}}) - \min_{\mathbf{R}}(E_{\mathrm{full}} - E_{\mathrm{lowrank}})
\end{equation}
over the geometries included in the corresponding set.

The results show a strong correlation between the truncation threshold imposed during decomposition of the training 2(t)RDMs and the resulting MAE of the interpolated energies, both on the training set and on the unseen test set. This supports the use of the two-body energy threshold as a practical and robust heuristic for selecting the compression rank at the training stage. The observed MAE lies slightly above the \(x=y\) line, which is consistent with the accumulation of small errors across the full set of 990 compressed 2(t)RDMs, but the decomposition threshold nevertheless remains a reliable predictor of the final interpolation accuracy. The NPE also decreases systematically as the threshold is tightened, although it remains somewhat larger than the MAE throughout. Importantly, the errors on the test geometries are only modestly larger than those on the training geometries, demonstrating that the compressed representation transfers well to interpolation at nearby unseen nuclear configurations. %Overall, this analysis establishes the energy-based truncation criterion as a stable and practically useful metric for controlling the compression rank within the eigenvector-continuation workflow.

Figure~\ref{fig:h28_hamthr} also reports the compression ratios of the training 2(t)RDMs relative to full rank (right-hand \(y\)-axis) for the same set of truncation thresholds. Although these data are derived from the same underlying RDM ensemble as in Fig.~\ref{fig:h28-distributions}, the values differ slightly because the present compression is performed after orthogonalization via Eq.~\ref{eq:ortho_rdm}. This orthogonalization within the training manifold appears to improve the overall compressibility for a given energy threshold, presumably because the projected RDMs are represented in an orthonormalized subspace basis. At the practically important threshold of \(10^{-3}\)~Ha, the overall compression across the full training 2(t)RDM set improves by approximately 10\% relative to the unorthogonalized case. It is also notable that the compression remains highly effective up to thresholds of roughly \(10^{-3}\)~Ha, whereas the gains become substantially less pronounced for tighter thresholds, consistent with the behavior already observed in Fig.~\ref{fig:scaling}. In particular, more than half of the full rank is required once the threshold is tightened to \(10^{-5}\)~Ha.
Overall, these results indicate that energy-based compression metrics made locally at the level of individual training 2(t)RDMs remain quantitatively predictive of the global interpolation error, and are a stable and practically useful criteria which is essential for controlling the compression rank and deploying the method in large-scale interpolation workflows.

%Notes regarding energy error, MAE closely follows x=y line whereas NPE is roughly twice MAE (signs of the errors mostly match, but there is a variation in predicted training energies such that spread around the mean; i.e. standard deviation, is larger than MAE.

%NPE being larger than MAE might be a result of more aggressive truncation on transition RDMs (since their energy is smaller than diagonal RDMs). A secondary, coarser threshold on eigenvalue/singular value magnitude across all (t)RDMs might be explored in the future to address this issue. More sophisticated truncation schemes can also be explored where truncation is determined on the accuracy of the continuation across all states, rather than single 2RDM contractions for energy.

\subsection{Efficient DMRG-based nonadiabatic molecular dynamics}
\label{sec:namd}

While the energy error provides an important diagnostic of the compressed representation, the most stringent test is its performance within a full nonadiabatic molecular dynamics (NAMD) workflow, where errors in energies, forces, and nonadiabatic couplings (NACs) accumulate over time and affect the sampled nuclear phase space~\cite{CrespoOtero18-chemicalreview}. We therefore now assess the impact of compressing the training 2(t)RDMs on dynamical, structural, and spectroscopic observables within the eigenvector continuation interpolation framework, having removed the principal memory bottleneck associated with storing and contracting the full set of training 2(t)RDMs. The underlying NAMD methodology, and its combination with wavefunction interpolation, were described in detail in our previous work~\cite{Atalar24-eigconNAMD-faraday}.

A key advantage of the interpolation scheme in this context is that analytic nuclear derivatives of the subspace Hamiltonian are readily available, even when such derivatives are not directly accessible from the underlying training solver. Using the expressions derived in Sec.~\ref{sec:inference}, the two-electron contributions to both the state-specific nuclear forces and the NACs can be evaluated efficiently from the compressed representation~\cite{Atalar24-eigconNAMD-faraday}, according to
\begin{align}
    \frac{\partial E_\mu}{\partial\mathbf{R}} &= \mathbf{C}_\mu^T \frac{\partial \mathcal{H}}{\partial\mathbf{R}} \mathbf{C}_\mu, \\
    d_{\mu\nu} &= \frac{1}{E_\nu-E_\mu}\,
    \mathbf{C}_\mu^T \frac{\partial \mathcal{H}}{\partial\mathbf{R}} \mathbf{C}_\nu.
\end{align}
These quantities determine, respectively, the classical force acting on the nuclei on a given adiabatic surface and the probability of stochastic hopping between adiabatic states in the surface-hopping dynamics.

\begin{figure*}[htb]
\centering
\includegraphics[width = 0.98\linewidth]{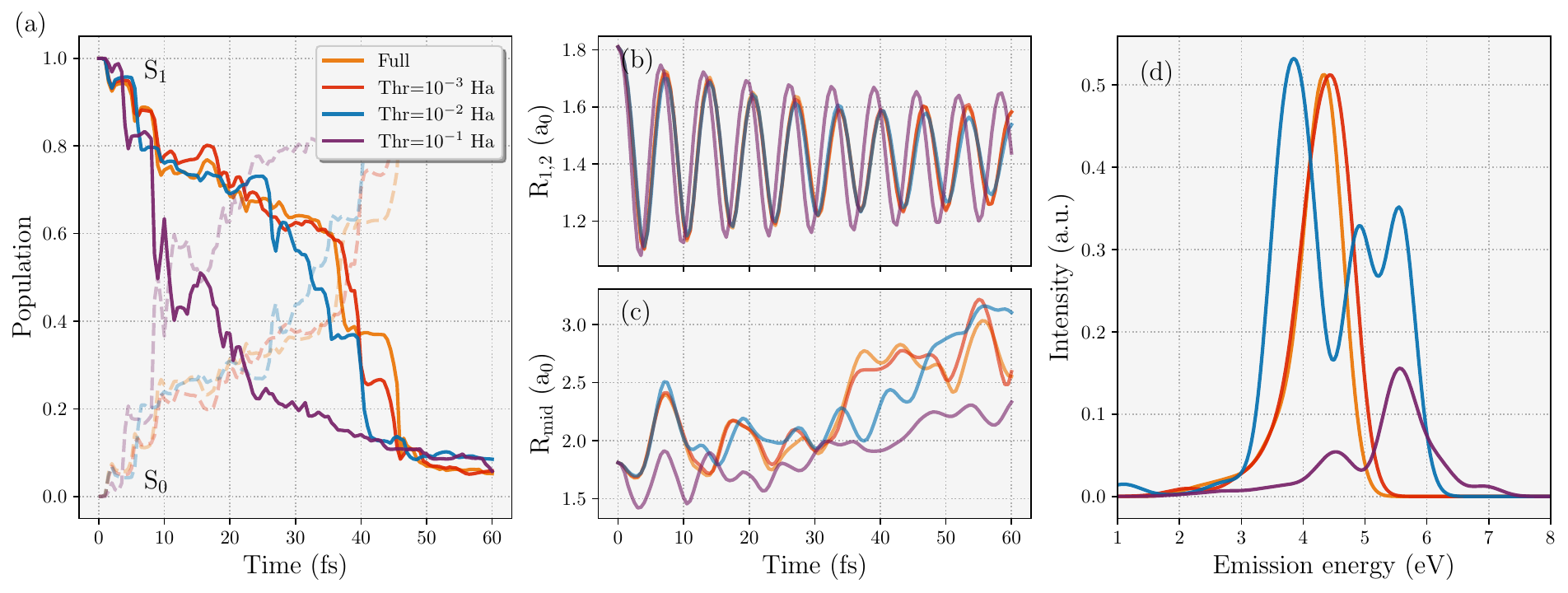}
\caption{Effect of the low-rank approximation on statistically averaged observables from a nonadiabatic molecular dynamics simulation of a photoexcited H$_{28}$ hydrogen chain in a STO-3G basis. (a) Population transfer from S$_{1}$ to S$_{0}$, (b) average distance between two terminal hydrogens, (c) average distance between two middle hydrogens, and (d) early fluorescence emission spectra (averaged for the 5 fs to 25 fs range) of the H$_{28}$ chain in minimal basis across several low-rank decompositions. All results are averaged over $\sim$100 trajectories, using a cc-pVDZ-RI auxiliary basis.}
\label{fig:namd}
\end{figure*}

We consider NAMD simulations for the H$_{28}$ chain following vertical photoexcitation to the first excited state, \(S_1\), whose subsequent relaxation dynamics were analyzed in Ref.~\onlinecite{Atalar24-eigconNAMD-faraday}. The energies, gradients, and NACs were inferred from the low-rank representations of the RDMs constructed from the 44 training states discussed in the previous sections. The dynamics were propagated using the fewest-switches surface-hopping (FSSH) algorithm~\cite{Tully90-FSSH} as implemented in \textsc{Newton-X}~\cite{NewtonX14,NewtonX22}. Our eigenvector-continuation implementation provided interpolated electronic energies, gradients, and NACs between all pairs of adiabatic states~\cite{Atalar24-eigconNAMD-faraday}. A nuclear time step of 0.5~fs was used, while the time-dependent electronic Schr\"odinger equation was propagated using a fifth-order multistep integrator due to Butcher~\cite{Butcher65-multistepODEintegrator} with 20 electronic substeps per nuclear step. Decoherence was treated using the simplified decay-of-mixing scheme~\cite{Granucci07-decoherence} with a decay parameter of 0.1~Ha.

To connect the dynamics to an experimentally relevant observable, time-resolved fluorescence emission spectra were also computed along each trajectory. The instantaneous emission intensity was evaluated as
\begin{equation}
I(\omega,t) =
P_{S_1}(t)\, |\mu_{01}(t)|^2 \, \omega^3 \,
\delta\!\left(\omega - \Delta E_{01}(t)\right),
\end{equation}
where \(P_{S_1}(t)\) is the excited-state population, \(\mu_{01}(t)\) is the transition dipole moment between the inferred \(S_1\) and \(S_0\) states, and \(\Delta E_{01}(t)\) is the instantaneous energy gap. The spectra were broadened with Gaussians of width 0.25~eV. Only portions of the trajectories prior to hopping to the ground state were included, since fluorescence cannot occur after nonradiative relaxation to \(S_0\). Emission intensities were averaged over the 5--25~fs time window to probe early-time fluorescence while excluding the initial vertical-excitation regime.

Figure~\ref{fig:namd} summarizes the effect of compressing the training 2(t)RDMs on a range of electronic, nuclear, and spectroscopic observables. By comparing trajectory ensembles generated with different truncation thresholds, we can assess how accumulated errors in interpolated energies, forces, and NACs affect the sampled nuclear phase space. In total, 100 trajectories with different random seeds were propagated for both the full interpolation and for each compressed model, allowing statistically averaged quantities to be compared.
We note that some trajectories were excluded from these averages, as they showed an energy drift throughout the simulations, with total energies diverging by more than 0.1~Ha compared to the value at $t=0$. In particular, this was observed for nine trajectories of the model with a low-rank truncation threshold of $10^{-1}$~Ha and five trajectories of the $10^{-2}$~Ha threshold. We attribute this to the combined effect of the low-rank approximation and density fitting, leading to steeper potential energy surfaces, as this did not occur for trajectories with the $10^{-3}$~Ha threshold. We expect that reducing the timestep would improve energy conservation in these trajectories, but we remove them entirely from the averages shown in Fig.~\ref{fig:namd} to allow a consistent comparison across low-rank approximations.
%\ghb{of what amount?} \ka{Added.}
%\ghb{This is unclear} \ka{See above paragraph}

%Fig.~\ref{fig:namd} compares the nonadiabatic molecular dynamics (NAMD) simulations obtained using different truncation levels of the RDMs. Electronic, nuclear, and spectroscopic observables are shown to assess the accuracy and robustness of the low-rank compression. It should be noted that the energies, forces and NACs in the low-rank inference is computed with JK builds using density-fitting (with the cc-pVDZ-RI auxillary basis) which introduces an additional (albeit small) error compared to full inference.

Figure~\ref{fig:namd}(a) shows the ensemble-averaged nonradiative electronic population decay back to the ground state following the initial \(S_0 \rightarrow S_1\) photoexcitation of the equidistant one-dimensional H$_{28}$ chain at the ground-state equilibrium geometry. This observable is particularly sensitive to the inferred NACs, since these govern the electronic transition probabilities. To probe the nuclear phase space sampled during the dynamics more directly, Figs.~\ref{fig:namd}(b--c) report two structural metrics: the distance between the terminal hydrogen atoms and the distance between the central hydrogen atoms. These provide a direct measure of the structural distortions driving the nonadiabatic dynamics. Figure~\ref{fig:namd}(d) presents the corresponding fluorescence emission spectra, which probe the effect of RDM truncation on excited-state energies and transition dipole moments, and therefore connect the compression error to experimentally accessible spectroscopic signatures.

The coarsest truncation threshold, \(10^{-1}\)~Ha, is clearly insufficient to reproduce the dynamics even qualitatively. Although the dimerization of the terminal hydrogen atoms in Fig.~\ref{fig:namd}(b) is captured at a broad qualitative level, the central hydrogen distance, which is more sensitive to the detailed trajectory ensemble, deviates rapidly from the full interpolation result. At this truncation level, both the population decay and the fluorescence spectrum are also qualitatively incorrect. Tightening the threshold to \(10^{-2}\)~Ha recovers the overall form of the population decay, but noticeable discrepancies remain. In particular, deviations in the central hydrogen distance emerge, with qualitative differences becoming apparent after roughly 30~fs, and the emission spectra exhibit artifacts such as additional double-peak structure, likely reflecting imperfect sampling of distinct regions of nuclear configuration space.

At a truncation threshold of \(10^{-3}\)~Ha, the compressed dynamics become quantitatively consistent with the full interpolation on the timescales relevant here. The population transfer is nearly indistinguishable from the uncompressed result, with only minor deviations appearing around 45~fs. The ensemble-averaged structural observables and the early-time fluorescence spectra are likewise essentially unchanged relative to the full calculation, up to the statistical uncertainty over the trajectories. This threshold is also consistent with the training/test error analysis of Sec.~\ref{sec:eigcon}, which identified \(10^{-3}\)~Ha as a point at which the rank reduction remain efficient while interpolation errors are quantitatively controlled. Although the present application remains limited to a relatively small system and basis, it nevertheless demonstrates that the low-rank compression is sufficiently accurate to preserve the key dynamical, structural, and spectroscopic signatures of the nonadiabatic relaxation. Overall, these results validate direct low-rank inference of energies, gradients, and NACs in this setting, and indicate that a truncation threshold of \(10^{-3}\)~Ha is adequate to achieve quantitative accuracy for NAMD simulations of this system. Future applications to larger systems will provide a more stringent test of how the required truncation threshold scales with basis size, state density, and the complexity of the underlying nonadiabatic landscape.

%% file: 6-conclusion.tex
\section{Conclusions \& Outlook}

In this work, we have investigated a low-rank compression framework for two-body and transition two-body reduced density matrices that is designed to preserve the fermionic wedge-product structure of two-body correlations under truncation. We also show how this form allows efficient extraction of observables directly from the compressed form via AO-based Coulomb and exchange builds. By coupling the Coulomb and exchange channels through a common set of low-rank factors, the decomposition provides a more compact representation of fermionic two-particle structure than any single-channel forms. The resulting compressed representation remains physically consistent under truncation, and provides the exact low-rank decomposition for mean-field states and compact active-space wavefunctions. For general correlated states, the memory scaling of two-electron (transition) density matrices is reduced from \(\mathcal{O}(M^4)\) to \(\mathcal{O}(rM^2)\), where the retained rank \(r\) was found to grow approximately linearly with system size at fixed absolute energy accuracy, consistent with the locality of irreducible two-body correlations. Given the size extensivity of the total energy, this further implies a system size independent rank for a given relative energy error, or precision per particle. We further demonstrated that inexpensive diagonal corrections in a local representation can restore physically important subsets of the tensor and yield additional compression gains. Together, these lead to substantial compression of the two-body description even at moderately sized systems, while maintaining millihartree-level accuracy in the inferred energies.

Overall, this scheme enables the compressed representation to be used as a practical surrogate for the original 2(t)RDMs in downstream workflows. A central motivating application was the recently introduced \emph{ab initio} eigenvector continuation interpolation framework, in which a manifold of training wavefunctions is used as a many-body basis to define a transferable subspace expansion across nuclear geometries. In that setting, the number of required 2(t)RDMs grows quadratically with the number of training points, while the storage of each object scales quartically with basis size, making these tensors a practical bottleneck of the method. By applying the present compression, we substantially reduced this overhead while retaining a controllable approximation to the downstream inference, together with a robust truncation metric for selecting the retained rank. This was applied to a realistic nonadiabatic molecular dynamics worflow, using systematically compressed near-exact DMRG training 2tRDMs to accumulate static and dynamical statistical observables for the photoexcitation and subsequent structural and electronic relaxation of a linear H$_{28}$ chain. In this setting, a target accuracy of 1~mHa was sufficient to recover statistically resolved observables quantitatively on the timescales relevant to the dynamics.

There are several natural directions for future work, as well as broader opportunities for this compression scheme in workflows beyond eigenvector continuation. Physical insight into correlated electronic structure and excitation character may be extracted from the decomposition metrics themselves~\cite{HeadGordon04-SVD-T2}, while the low-rank form could also be explored as a direct optimization variable. Within machine-learning workflows, such reduced-rank representations may lower resource requirements while imposing more physically meaningful inductive biases on the model~\cite{b2026machinelearningtwoelectronreduced}. This may also be valuable in quantum-classical settings, where incomplete or noisy two-body data must be regularized and the compressed form used for reduced density matrix reconstruction and noise filtering~\cite{Chan23-RDM-completion,doi:10.1021/acs.jpclett.6c00296}. Finally, the scheme could be generalized to higher-order reduced density matrices, or to dynamical and nonequilibrium two-body variables in which time- or frequency-domain objects can be analogously compressed. More broadly, the present results suggest that low-rank, structure-preserving two-body variables can serve not only as formally complete two-body descriptors, but also as practically compressible and transferable intermediate representations for correlated electronic structure.

%% file: acknowledgements.tex
\section*{Acknowledgements}

We wish to thank Yannic Rath for helpful discussions on this work. G.H.B and K.A. gratefully acknowledge support from the Air Force Office of Scientific Research under award number FA8655-22-1-7011.
H.G.A.B. is supported by a Royal Society University Research Fellowship (URF\textbackslash{}R0\textbackslash{}241299) at University College London.

\ifjcp
    %\section*{Author declarations}
    %\subsection*{Conflict of Interest}
    %The authors have no conflicts to disclose.

    %\subsection*{Author Contributions}

%\noindent \textbf{K.A.}: Conceptualization (equal); Data curation (lead); Formal analysis (equal); Investigation (equal); Methodology (equal); Software (equal); Validation (equal); Visualization (lead); Writing -- original draft (equal); Writing -- review \& editing (equal). \textbf{H.G.A.B.}: Methodology (supporting); Writing -- original draft (supporting); Writing -- review \& editing (equal). \textbf{A.G.}: Methodology (supporting); Writing -- review \& editing (equal). \textbf{G.H.B.}: Conceptualization (equal); Formal analysis (equal); Funding acquisition (lead); Investigation (equal); Methodology (equal); Project administration (lead); Resources (lead); Software (equal); Supervision (lead); Validation (equal); Writing -- original draft (equal); Writing -- review \& editing (equal).

\fi

\section*{Data availability}

Open-source code for generating the results of low-rank compression of 2-tRDMs, inference in the low-rank representation and ab initio eigenvector continuation, along with examples, can be found at https://github.com/BoothGroup/evcont. The raw data that support the findings of this study are available from the corresponding authors upon reasonable request.

%% file: appendix.tex
\appendix
\section*{Appendices}

\section{Relaxed amplitude low-rank fit}
\label{app:relax}

\begin{figure}[h]
\centering
  \includegraphics[width=0.95\linewidth]{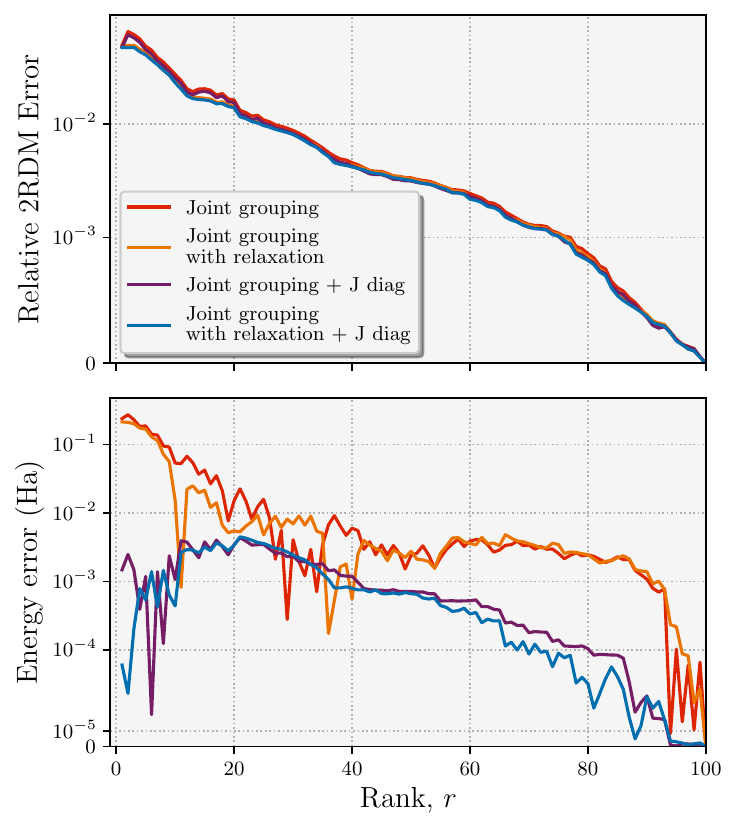}
  \caption{Low-rank joint decomposition of the FCI two-body reduced density matrix of H$_{10}$ in a minimal STO-6G basis at $1.5$a$_0$ interatomic separation, with and without the Coulomb ($J$) diagonal correction, and with and without further linear amplitude relaxation. Upper plot characterizes the relative mean absolute error over all 2RDM elements, while the lower plot characterizes the absolute two-electron energy error from the reconstructed 2RDM.}
  \label{fig:fci-decomp-relax}
\end{figure}

In Fig.~\ref{fig:fci-decomp-relax} we show results from the relaxation of the linear coefficients in the low-rank joint expansion described in Sec.~\ref{sec:relax}. For the same system and setup as Fig.~\ref{fig:fci-decomp}, we also include the amplitude relaxation, both with and without the diagonal corrections, and observe the impact on the norm error of the reconstructed 2RDM and two-body energy. Both error metrics show modest improvements from the unrelaxed amplitude fidelities across all ranks of the decomposition for this system.

\section{Low-rank form of transition 2RDM between single Slater determinants}
\label{app:trdm}

In this appendix we derive a compact and numerically stable representation of the transition two-body reduced density matrix (2tRDM) between two single Slater determinants constructed from nonorthogonal orbital sets, and extend the resulting expression to the case in which the determinant overlap may vanish. To streamline the notation, we adopt the co-/contravariant labelling convention of Ref.~\onlinecite{10.1063/1.475423}, with subscripts and superscripts denoting covariant and contravariant indices, respectively.

%\subsubsection*{Biorthogonalisation of the occupied spaces}

Consider two $N$-electron Slater determinants $|\Psi_a\rangle$ and $|\Psi_b\rangle$ constructed from occupied spin orbital sets $\{\,{}^{a}\phi_{i\sigma}\,\}$ and $\{\,{}^{b}\phi_{j\sigma}\,\}$, where $i \sigma$ is a combined spatial-spin orbital index with $\sigma \in \{\uparrow, \downarrow\}$.
First, we biorthogonalize the occupied spaces of the two states by defining the occupied–occupied overlap matrix
\begin{equation}
{}^{ab}S_{i\sigma,j\tau} = \langle {}^{a}\phi_{i\sigma} \,|\, {}^{b}\phi_{j\tau}\rangle .
\end{equation}
A singular value decomposition (L\"{o}wdin pairing)
\begin{equation}
{}^{ab}\bm{S} = \bm{U}\, \mathrm{diag}(s_{i\sigma})\, \bm{V}^\dagger
\end{equation}
defines the biorthogonal occupied orbitals\cite{10.1098/rspa.1961.0175,10.1063/1.1712221}
\begin{equation}
|{}^{a}\tilde{\phi}_{i\sigma}\rangle = \sum_{j\tau} |{}^{a}\phi_{j\tau}\rangle U_{j\tau,i\sigma}, 
\quad
|{}^{b}\tilde{\phi}_{i\sigma}\rangle = \sum_{j\tau} |{}^{b}\phi_{j\tau}\rangle V_{j\tau,i\sigma},
\end{equation}
such that the spin orbitals are pairwise biorthogonal, and the determinant overlap is given by the product of the singular values, 
\begin{equation}
\langle {}^{a}\tilde{\phi}_{i\sigma} \,|\, {}^{b}\tilde{\phi}_{j\tau} \rangle = \delta_{ij} \delta_{\sigma\tau}s_{i\sigma},
\quad
{}^{ab}S = \langle \Psi_a | \Psi_b \rangle = \prod_{i\sigma} s_{i\sigma} .
\end{equation}
%
%\subsubsection*{Orbital-resolved one-body transition matrices}
%In what follows, we index the spin orbitals using the explicit notation $p \rightarrow p\sigma$, where $\sigma \in \{\uparrow, \downarrow\}$ denotes the orbital spin.
We can now define the orbital-resolved one-body transition matrices, as
\begin{equation}
P_{m\sigma}^{ij} = ({}^{b}\tilde{C})^{i \cdot}_{\cdot, m\sigma}\, ({}^{a}\tilde{C}^{*})_{m\sigma,\cdot}^{\cdot j},
\end{equation}
such that each $\bm{P}_{m\sigma}$ corresponds to a single occupied spin-orbital pair.
In the biorthogonal basis, generalized Wick contractions factor through these rank-one contributions~\cite{10.1063/5.0045442,10.1063/5.0122094,10.1063/5.0156124,10.1063/5.0246790}, yielding the exact 2tRDM
\begin{equation}
{}^{ab}\Gamma_{ijkl}
=
\sum_{mn}\sum_{\sigma \tau} \epsilon_{m\sigma,n\tau}
\left(
P_{m\sigma}^{ij} P_{n\tau}^{kl}
-
\delta_{\sigma \tau}
P_{m\sigma}^{il} P_{n\tau}^{kj}
\right).
\label{eq:app_general_double_sum}
\end{equation}
%\ghb{There seems to be an inconsistency here with the definition in Eq.~\ref{eq:hf-2rdm}. Appropriate product should be $(ij)(kl)-(ik)(lj)$? Also, should we not spin-integrate for the appropriate spatial orbital factors? This convention discrepancy will likely propagate down into Eq.\ref{eq:app_partitioned} and Eq.\ref{eq:app_final}, and indeed Eq.\ref{eq:factorised_tRDM}? Haven't changed - Check for consistency, along with the index ordering consistency for the M, W etc factors later.} \ka{Fixed.}
The weights are defined as
\begin{equation}
\epsilon_{m\sigma,n\tau} = \prod_{p\gamma\neq m\sigma,n\tau} s_{p\gamma} 
\end{equation}
where the index $p$ runs over all occupied spin orbitals in the biorthogonal basis not equal to $m\sigma$ or $n\tau$.
Note that the diagonal entries $\epsilon_{m\sigma,m\sigma}$ can be arbitrarily defined since
$P_{m\sigma}^{ij} P_{m\sigma}^{kl} - 
P_{m\sigma}^{il} P_{m\sigma}^{kj} = 0$.

The factorization in Eq.~\eqref{eq:app_general_double_sum} is valid and numerically robust for any value of the many-body overlap $\braket{\Psi_a|\Psi_b}$, including the zero-overlap case $\braket{\Psi_a|\Psi_b}=0$\cite{10.1063/5.0156124}.
Diagonalizing the real symmetric $\epsilon_{m\sigma,n\tau}$ tensor as
\begin{equation}
\epsilon_{m\sigma,n\tau} = \sum_{x} Q_{m\sigma,x} \kappa_{x} Q_{n\tau,x}
\end{equation}
leads to further factorization of the 2tRDM as
\begin{equation}
{}^{ab}\Gamma_{ijkl}
=
\sum_{x} \sum_{\sigma \tau}
\kappa_{x}
\left(
P_{x\sigma}^{ij} P_{x\tau}^{kl}
-
\delta_{\sigma \tau}
P_{x\sigma}^{il} P_{x\tau}^{kj}
\right),
\label{eq:full_factor}
\end{equation}
where now
\begin{equation}
    P_{x\sigma}^{ij} = \sum_{m} P_{m\sigma}^{ij}\;  Q_{m\sigma,x} .
\end{equation}
Equation~\eqref{eq:full_factor} demonstrates that the same-spin components can be exactly represented using a wedge-product decomposition as
\begin{equation}
{}^{ab}\Gamma_{ijkl}^{\text{same}}
= \sum_{x\sigma} \kappa_{x}
\left( P_{x\sigma}^{ij} P_{x\sigma}^{kl} - P_{x\sigma}^{il} P_{x\sigma}^{kj} \right),
\end{equation}
while the different-spin component adopts the form of a Coulomb decomposition
\begin{equation}
{}^{ab}\Gamma_{ijkl}^{\text{diff}}
= \sum_{x} \kappa_{x}
\left(P_{x +}^{ij} P_{x +}^{kl} - P_{x -}^{ij} P_{x -}^{kl} \right),
\end{equation}
where $P_{x \pm}^{ij} = \frac{1}{\sqrt{2}} (P_{x \uparrow}^{ij} \pm P_{x \downarrow}^{ij})$.
In the spin-symmetric case 
\begin{equation}
P_{x\uparrow}^{ij} =  P_{x\downarrow}^{ij} = \frac{1}{2}P_x^{ij}
\end{equation} 
(e.g., two closed-shell determinants),
we recover the  joint decomposition of the full 2tRDM as
\begin{equation}
{}^{ab}\Gamma_{ijkl}
= \sum_{x} \kappa_{x}
\left( P_{x}^{ij} P_{x}^{kl} - \frac{1}{2}P_{x}^{il} P_{x}^{kj} \right).
\end{equation}
These expressions justify the use of a wedge product decomposition and a Coulomb decomposition for the same-spin and different-spin components of the transition 2RDM, respectively, for states which can be decomposed into a formally low-rank number of determinants.

We now connect this result to the generalized Slater--Condon rules, showing that the same-spin and different-spin transition 2RDMs obtained from these rules also satisfy a wedge-product and Coulomb decomposition, respectively.
Following Chen and Scuseria\cite{10.1063/5.0156124}, we partition the biorthogonalized occupied orbitals into two sets: strongly overlapping orbitals where $|s_{i\sigma}| > a$, and weakly overlapping orbitals with $|s_{v\sigma}| \le a$, for some small threshold, $a$.
Here, \(v\sigma,t\tau\) label orbitals in the weak-overlap subset \(|s_{v\sigma}|\le a\), while the index \(p\gamma\) indexes all occupied biorthogonal orbitals. 
We can then define the spin-resolved co-weighted transition density matrix
\begin{equation}
W_\sigma^{ij} = \sum_{|s_{m\sigma}|>a} \frac{1}{s_{m\sigma}} P_{m\sigma}^{ij},
\end{equation}
with the total density \(W^{ij} = W_\uparrow^{ij} + W_\downarrow^{ij}\).
%and the reduced weights
%\begin{equation}
%\epsilon_{v\sigma} = \prod_{p\gamma\neq v\sigma} s_{p\gamma},
%\qquad
%\epsilon_{v\sigma,t\tau} = \prod_{p\gamma\neq v\sigma,t\tau} s_{p\gamma}.
%\end{equation}
The summation in \(W_\sigma^{ij}\) runs over the well-overlapping subset \(|s_{m\sigma}|>a\).
%\ghb{Again, clarify $p\neq v,t$ and $p\neq v$ orbital sums.} \ka{Done below.}
%Here and below, \(v\sigma,t\tau\) label orbitals in the weak-overlap subset \(|s_{v\sigma}|\le a\), while the index \(p\gamma\) in the products runs over all occupied biorthogonal orbitals. 
%The sums entering \(W_\sigma^{ij}\) run over the complementary well-overlapping subset \(|s_{m\sigma}|>a\).
The same-spin transition 2RDM is then partitioned as
\begin{equation}
\begin{aligned}
&{}^{ab}\Gamma^{\text{same}}_{ijkl}
=
{}^{ab}S \sum_{\sigma}\left( W_\sigma^{ij} W_\sigma^{kl} - W_\sigma^{il} W_\sigma^{kj} \right)
\\
&\quad + \sum_{v\sigma} \epsilon_{v\sigma}
\left(
P_{v\sigma}^{ij} W_\sigma^{kl}
+ W_\sigma^{ij} P_{v\sigma}^{kl} 
- P_{v\sigma}^{il} W_\sigma^{kj}
- W_\sigma^{il} P_{v\sigma}^{kj}
\right)
\\
&\quad
+ \sum_{v\sigma, t\sigma} \epsilon_{v\sigma,t\sigma}
\left(
P_{v\sigma}^{ij} P_{t\sigma}^{kl} - P_{v\sigma}^{il} P_{t\sigma}^{kj}
\right).
\end{aligned}
\label{eq:app_partitioned}
\end{equation}
%We can introduce this expression to 
%\begin{equation}
%M_{\sigma}^{ij} = \sum_v \epsilon_{v\sigma} P_{v\sigma}^{ij},
%\end{equation}
%to further simplify the transition 2RDM to the form
%\begin{equation}
%\begin{aligned}
%&{}^{ab}\Gamma^{\text{same}}_{ijkl}
%=
%{}^{ab}S \sum_{\sigma}\left( W_\sigma^{ij} W_\sigma^{kl} - W_\sigma^{il} W_\sigma^{kj} \right)
%\\
%&\quad + \sum_{\sigma} 
%\left(
%M_{\sigma}^{ij} W_\sigma^{kl}
%+ W_\sigma^{ij} M_{\sigma}^{kl} 
%- M_{\sigma}^{il} W_\sigma^{kj}
%- W_\sigma^{il} M_{\sigma}^{kj}
%\right)
%\\
%&\quad
%+ \sum_{v\sigma, t\sigma} \epsilon_{v\sigma,t\sigma}
%\left(
%P_{v\sigma}^{ij} P_{t\sigma}^{kl} - P_{v\sigma}^{il} P_{t\sigma}^{kj}
%\right).
%\end{aligned}
%\label{eq:app_partitioned_more}
%\end{equation}
This expression provides a numerically robust form of the corresponding generalised Slater--Condon rules for two arbitrary Slater determinants.
Similarly, the different-spin transition 2RDM is given by 
\begin{equation}
\begin{aligned}
{}^{ab}\Gamma^{\text{diff}}_{ijkl}
=
{}^{ab}S \sum_{\sigma\neq \tau} \Big[ &W_\sigma^{ij} W_\tau^{kl} 
\\
+ &\sum_{v} \epsilon_{v\sigma}
\left(
P_{v\sigma}^{ij} W_\tau^{kl}
+ W_\tau^{ij} P_{v\sigma}^{kl} 
\right)
\\
+ &\sum_{vt} \epsilon_{v\sigma,t\tau}
P_{v\sigma}^{ij} P_{t\tau}^{kl} \Big].
\end{aligned}
\label{eq:app_partitioned_diff}
\end{equation}

We now consider the different possible cases depending on how many biorthogonal orbital pairs have an overlap below the threshold $a$.
In what follows, we make use of the corresponding expression for the spin-resolved 1tRDM given by
\begin{equation}
{}^{ab}\gamma^{ij}_\sigma = \sum_{m} \epsilon_{m\sigma} P_{m\sigma}^{ij} = 
{}^{ab}S W^{ij}_\sigma + \sum_{v \sigma} P_{v \sigma}^{ij}
\end{equation}
with ${}^{ab}\gamma^{ij} = {}^{ab}\gamma^{ij}_\uparrow + {}^{ab}\gamma^{ij}_\downarrow$.

\underline{No zero overlaps:}
If all $\left|s_{v\sigma}\right| > a$, then $^{ab}S\neq0$ and the set of weakly overlapping orbitals is empty. 
In this case, the same-spin 2tRDM requires a rank-1 wedge-product
\begin{equation}
{}^{ab}\Gamma^{\text{same}}_{ijkl}
=
{}^{ab}S \sum_{\sigma}\left( W_\sigma^{ij} W_\sigma^{kl} - W_\sigma^{il} W_\sigma^{kj} \right).
\end{equation}
The analogous expression for the different-spin 2tRDM is
\begin{equation}
{}^{ab}\Gamma^{\text{diff}}_{ijkl} = {}^{ab}S \sum_{\sigma \neq \tau}  W_\sigma^{ij} W_\tau^{kl}
 = W_+^{ij} W_+^{kl} - W_-^{ij} W_-^{kl},
\end{equation}
which adopts the form of a rank-2 Coulomb decomposition.
These two expressions can be combined to give the overall 2tRDM 
\begin{equation}
{}^{ab}\Gamma_{ijkl} = \frac{1}{{}^{ab}S} ({}^{ab}\gamma^{ij}\, {}^{ab}\gamma^{kl} - \sum_\sigma {}^{ab}\gamma^{il}_\sigma\, {}^{ab}\gamma^{kj}_{\sigma} ).
\label{eq:2RDM_nozero}
\end{equation}
In the spin-symmetric case, where ${}^{ab}\gamma^{ij}_{\uparrow} = {}^{ab}\gamma^{ij}_{\downarrow} = \frac{1}{2} {}^{ab}\gamma^{ij}$, Eq.~\eqref{eq:2RDM_nozero} reduces to the rank-1 wedge-product structure given in Eq.~\eqref{eq:hf-2rdm}.

\underline{One zero overlap:}
Next, we consider the case where the biorthogonalization yields one zero-overlap orbital pair with index $v \sigma$.
In this situation, we obtain ${}^{ab}S = 0$ and only one $\epsilon_{v \sigma} \neq 0$.
The non-zero contribution to the same-spin 2tRDM is then 
\begin{equation}
{}^{ab}\Gamma^{\text{same}}_{ijkl}
=
{}^{ab}\gamma^{ij}\, W_\sigma^{kl}
+ W_\sigma^{ij}\, {}^{ab}\gamma^{kl}
- {}^{ab}\gamma^{il}\, W_\sigma^{kj}
- W_\sigma^{il}\, {}^{ab}\gamma^{kj}
\label{eq:2RDMsame_1zero}
\end{equation}
where we have inserted the definition of the 1tRDM in this case as ${}^{ab}\gamma^{ij} = \epsilon_{v \sigma} P_{v\sigma}^{ij}$.
This expression is identical to the corresponding result from the generalized Slater--Condon rules.\cite{10.1063/5.0045442,10.1063/5.0122094}
Equation~\eqref{eq:2RDMsame_1zero} can be factorized into a rank-2 wedge-product structure as
\begin{equation}
{}^{ab}\Gamma^{\text{same}}_{ijkl}
= \frac{1}{2} \left[
\left(
P^{ij}_+ P^{kl}_+ - P^{il}_+ P^{kj}_+
\right)
-
\left(
P^{ij}_- P^{kl}_- - P^{il}_- P^{kj}_- 
\right)\right]
\end{equation}
where the basis vectors are $P^{ij}_\pm = {}^{ab}\gamma^{ij} \pm W_\sigma^{ij}$ and $\sigma$ corresponds to the spin of the zero-overlap orbital pair.
The different-spin 2tRDM is given by a rank-2 Coulomb decomposition as 
\begin{equation}
{}^{ab}\Gamma^{\text{diff}}_{ijkl}
= 
\frac{1}{2} \left( Q^{ij}_+ Q^{kl}_+ - Q^{ij}_- Q^{kl}_- \right)
\end{equation}
where now $Q^{ij}_\pm =  {}^{ab}\gamma^{ij} \pm W_\tau^{ij}$ with $\tau \neq \sigma$.
Under the approximation $W^{ij}_{\uparrow} \approx W^{ij}_{\downarrow}$, which can be assumed
for spin-restricted orbitals when the number of non-zero overlapping orbitals is large,
then $Q_{\pm}^{ij} \approx P^{ij}_\pm$ and the full 2tRDM adopts an approximate rank-2 joint decomposition as
\begin{equation}
{}^{ab}\Gamma_{ijkl}
\approx \big(
P^{ij}_+ P^{kl}_+ - \frac{1}{2}P^{il}_+ P^{kj}_+
\big)
-
\big(
P^{ij}_- P^{kl}_- - \frac{1}{2}P^{il}_- P^{kj}_- 
\big).
\end{equation}

\underline{Two zero overlaps:}
Finally, we consider the case where two orbitals have a vanishing biorthogonal overlap, corresponding to spin-orbital indices $v\sigma$ and $t \tau$.
In this situation, ${}^{ab}S = 0$ and all $\epsilon_{p\gamma} = 0$.
The only non-zero contribution to the 2tRDM is 
\begin{equation}
{}^{ab}\Gamma_{ijkl} = \epsilon_{v\sigma,t\sigma}
( P_{v\sigma}^{ij} P_{t\tau}^{kl} - \delta_{\sigma \tau} P_{v\sigma}^{il} P_{t\tau}^{kj} ),
\end{equation}
which can be expressed as the rank-2 wedge-product 
\begin{equation}
\begin{aligned}
{}^{ab}\Gamma_{ijkl} = \frac{\epsilon_{v\sigma,t\sigma}}{2}
\Big[
&(
P^{ij}_+ P^{kl}_+ -  \delta_{\sigma \tau} P^{il}_+ P^{kj}_+
)
\\
-&(
P^{ij}_- P^{kl}_- - \delta_{\sigma \tau} P^{il}_- P^{kj}_-
)
\Big]
\end{aligned}
\end{equation}
where $P^{ij}_\pm = P_{v\sigma}^{ij} \pm P_{t\tau}^{ij}$.

\section{Computations with low-rank representations}
\label{app:inference}

We expand on the expressions derived in Sec.~\ref{sec:inference} with further details on the derivation of expectation values from the low rank form, including the 
%cases of an SVD representation of 2RDMs in Coulomb grouping, 
diagonal corrections, and Pulay-like contributions to the energy gradients in the joint decomposition.

\subsubsection*{Orbital derivative contributions to nuclear gradients}

%\ghb{Are we going to insert back the $\varepsilon^{(\alpha)}$ eigenvalues explicitly into the form, derived from Eq.~\ref{eq:coulombvec_jointdecomp}, rather than \ref{eq:joint_decomp}?}
%\ka{Done}

The orbital derivative (Pulay) contributions arise from the nuclear derivative of the orbital transformation appearing in the four–index transformation of the two–electron integrals. The contribution to the gradient from these terms can be written as
\begin{equation}
\nabla_{\mathbf R} E^{(\mathrm{pulay})}
=
\frac{1}{2} \,
\Gamma_{ijkl}
\nabla_{\mathbf R} \left[ Z_{wi} Z_{xj} Z_{yk} Z_{zl}
(wx|yz) \right].
\end{equation}

\noindent Expanding the derivative produces four contributions,
\begin{align}
\nabla_{\mathbf R} E^{(\mathrm{pulay})}
=
\frac{1}{2}\Gamma_{ijkl}
\Big[
&(\nabla_{\mathbf R} Z_{wi}) Z_{xj} Z_{yk} Z_{zl}
\nonumber\\
&+
Z_{wi} (\nabla_{\mathbf R} Z_{xj}) Z_{yk} Z_{zl}
\nonumber\\
&+
Z_{wi} Z_{xj} (\nabla_{\mathbf R} Z_{yk}) Z_{zl}
\nonumber\\
&+
Z_{wi} Z_{xj} Z_{yk} (\nabla_{\mathbf R} Z_{zl})
\Big]
(wx|yz). \label{eq:fourderivs}
\end{align}

Substituting the joint decomposition of the 2RDM defined in Eq.~\ref{eq:joint_decomp}, the first derivative contribution can be written as
\begin{align}
(\nabla_{\mathbf R} Z_{wi}) ~ \varepsilon^{(\alpha)}
\Big[
&Z_{xj} v_{ij}^{(\alpha)}
\sum_{yz} (wx|yz) v_{yz}^{(\alpha)}
\nonumber \\
&-
\frac{1}{2}
Z_{zl} v_{il}^{(\alpha)}
\sum_{yz} (wx|yz) v_{yx}^{(\alpha)}
\Big].
\end{align}
The second term in the Eq.~\ref{eq:fourderivs} can also be written similarly and relabelled as
\begin{align}
(\nabla_{\mathbf R} Z_{wi})~ \varepsilon^{(\alpha)}
\Big[
&Z_{xj} \left(v_{ij}^{(\alpha)}\right)^T
\sum_{yz} (wx|yz) v_{yz}^{(\alpha)}
\nonumber \\
&-
\frac{1}{2}
Z_{zl} \left(v_{il}^{(\alpha)}\right)^T
\sum_{yx} (wx|yz) v_{yx}^{(\alpha)}
\Big].
\end{align}

Using the Coulomb and exchange builds defined in
Eq.~\ref{eq:jkbuilds}, these contractions can be combined into the
Fock-like intermediates
\begin{equation}
F_{wx}^{(\alpha)} = J_{wx} \left[ v^{(\alpha)} \right] - \tfrac{1}{2} K_{wx} \left[ v^{(\alpha)} \right] .
\end{equation}
%\ghb{I would make explicit the vector which is used as the `density' in the construction of these J and K matrices.} \ka{Better? Exchange build is defined slightly differently to allow the same matrix as input but they are equivalent. i.e. $\sum_{yx} (wx|yz) v_{yx}^{(\alpha)} = \sum_{yz} (wz|yx) v_{yz}^{(\alpha)}$ as defined in Eq.~\ref{eq:jkbuilds}. it is also why I choose to omit the indices of the input matrix as the equation explicitly contracts in its given indices.}
The remaining derivative contributions are equivalent after relabelling of indices, and the symmetry of the two–electron integrals allows the resulting contractions to be expressed using the same intermediates. Collecting the terms yields
\begin{equation}
\nabla_\mathbf{R} E^\mathrm{(pulay)} =
\left(\nabla_\mathbf{R}Z_{wi} \right)
\, Z_{xj} ~ \varepsilon^{(\alpha)} \,
\left[
v_{ij}^{(\alpha)} F_{wx}^{(\alpha)}
+
v_{ji}^{(\alpha)} F_{xw}^{(\alpha)}
\right].
\end{equation}

\subsubsection*{Diagonal corrections for energy evaluation} 

In this section we focus on efficient evaluation of the $D^1_{ij}$ diagonal correction as defined in Eq.~\ref{eq:diagJ}, as this is the most important one, used in the practical workflows, while the other diagonal corrections can also be constructed analogously. The correction to the 2RDM can be written as
\begin{equation}
\Delta\Gamma^{(1)}_{ijkl}
=
D^1_{\mu\nu}\,\delta_{\mu i}\delta_{\mu j}\delta_{\nu k}\delta_{\nu l}.
\end{equation}

Working with a Cholesky or density fitting approximation, where we can write $(ij|kl)=L_{ij,\alpha} L_{kl,\alpha}$, the electronic energy correction due to the diagonal term can be written as:
\begin{equation}
\Delta E_{ee}^{(1)}
=
\frac{1}{2}
D^1_{ij}\, L_{ii,\alpha} L_{jj,\alpha}.
\end{equation}

Constructing the contraction with efficient Coulomb builds ($J_{wx} = (wx|yz) v_{yz}$) is more difficult, as the diagonals are extracted in the SAO basis. To convert to the AO basis, the intermediate:
\begin{equation}
T_{wx,j}^{(1)} = D^{(1)}_{ij} Z_{wi} Z_{xi}
\end{equation}
can be constructed such that the two corresponding AO indices $w,x$ are obtained. The two-electron energy correction can then be computed as:
\begin{equation}
\Delta E_{ee}^{(1)}
=
\frac{1}{2}
Z_{yj} Z_{zj}\; J_{yz,j}[T_{wx,j}^{(1)}].
\end{equation}

Note that this requires $\mathcal{O}(M)$ Coulomb builds, as a computation is performed for each $j$ index, leading to an overall formal $\mathcal{O}(M^4)$ scaling. The cases for the other two diagonal corrections ($D_{ij}^{(2)}$ and $D_{ij}^{(3)}$) as defined in Eq.~\ref{eq:diagK1} and~\ref{eq:diagK2} are similar, but due to the orbital indices stored, they require $\mathcal{O}(M)$ exchange builds, leading to a formal $\mathcal{O}(M^5)$ scaling. In large-scale schemes, the locality of the SAO basis could likely be exploited to reduce these costs significantly, but this was not explored in this work. However, since the additional benefit from the $K$-diagonals is small compared to the $J$-diagonals, they are usually not included in the low-rank representation unless explicitly stated, ensuring that only Coulomb matrix builds are required.

\subsubsection*{Diagonal corrections for nuclear gradients}

Similar contractions can be performed for the nuclear gradients, involving both the two-electron integrals and Pulay terms as defined in Eq.~\ref{eq:gradE}. The derivation is shown for $D^{(1)}_{ij}$ using gradient Coulomb builds; corresponding expressions can be derived for $D^{(2,3)}_{ij}$ using gradient exchange builds. Initially, the 2RDM correction in the AO basis can be written as:
\begin{equation}
\Delta\Gamma^{(1)}_{wxyz}
=
D^1_{ij}\,Z_{wi}Z_{xi}Z_{yj}Z_{zj}.
\end{equation}

The electron repulsion integral gradient contribution as defined in Eq.~\ref{eq:eri_grad} becomes
\begin{equation}
\nabla_{\mathbf{R}} E^{(\mathrm{eri})}
=
4\, Z_{yj}Z_{zj}
\left(
\nabla_{\mathbf{R}} J_{yz,j}[T_{wx,j}^{(1)}]
\right).
\end{equation}

The Pulay terms can be derived similarly:
\begin{equation}
\nabla_{\mathbf{R}} E^{(\mathrm{pulay})}
=
2\, (\nabla_{\mathbf{R}} Z_{yj}) Z_{zj}
\left[
J_{yz,j}[T_{wx,j}^{(1)}]
+
J_{zy,j}[T_{wx,j}^{(1)}]
\right],
\end{equation}
where the Coulomb builds $J$ can be reused from the energy contraction.

For the $K$-diagonal corrections, the ERI gradient term becomes:
\begin{align}
\nabla_{\mathbf{R}} E^{(\mathrm{eri})}
=
2\, Z_{xj}Z_{zj} \Bigg[
&\nabla_{\mathbf{R}} K_{xz,j}\!\left[D^{(2,3)}_{ij} Z_{wi} Z_{yi}\right] \nonumber \\
&+
\nabla_{\mathbf{R}} K_{xz,j}\!\left[\left(D^{(2,3)}_{ij}\right)^T Z_{wi} Z_{yi}\right]
\Bigg],
\end{align}
while the Pulay term is computed as:
\begin{align}
\nabla_{\mathbf{R}} E^{(\mathrm{pulay})}
=
2\, (\nabla_{\mathbf{R}} Z_{xj}) Z_{zj} \Bigg[
&K_{xz,j}\!\left[D^{(2,3)}_{ij} Z_{wi} Z_{yi}\right] \nonumber \\
&+
K_{xz,j}\!\left[\left(D^{(2,3)}_{ij}\right)^T Z_{wi} Z_{yi}\right]
\Bigg].
\end{align}

Overall, these contractions scale similarly to the energy contractions, where $\mathcal{O}(M)$ Coulomb and gradient Coulomb builds are used for the $J$-diagonal corrections, whereas $\mathcal{O}(M)$ exchange and gradient exchange builds are required for the $K$-diagonal terms.
%\subsection{Further}

%% file: switchable_main.bbl
%aipnum4-2.bst 2019-01-14 (MD) hand-edited version of apsrev4-1.bst
\begin{thebibliography}{66}%
\makeatletter
\providecommand \@ifxundefined [1]{%
 \@ifx{#1\undefined}
}%
\providecommand \@ifnum [1]{%
 \ifnum #1\expandafter \@firstoftwo
 \else \expandafter \@secondoftwo
 \fi
}%
\providecommand \@ifx [1]{%
 \ifx #1\expandafter \@firstoftwo
 \else \expandafter \@secondoftwo
 \fi
}%
\providecommand \natexlab [1]{#1}%
\providecommand \enquote  [1]{``#1''}%
\providecommand \bibnamefont  [1]{#1}%
\providecommand \bibfnamefont [1]{#1}%
\providecommand \citenamefont [1]{#1}%
\providecommand \href@noop [0]{\@secondoftwo}%
\providecommand \href [0]{\begingroup \@sanitize@url \@href}%
\providecommand \@href[1]{\@@startlink{#1}\@@href}%
\providecommand \@@href[1]{\endgroup#1\@@endlink}%
\providecommand \@sanitize@url [0]{\catcode `\\12\catcode `\$12\catcode
  `\&12\catcode `\#12\catcode `\^12\catcode `\_12\catcode `\%12\relax}%
\providecommand \@@startlink[1]{}%
\providecommand \@@endlink[0]{}%
\providecommand \url  [0]{\begingroup\@sanitize@url \@url }%
\providecommand \@url [1]{\endgroup\@href {#1}{\urlprefix }}%
\providecommand \urlprefix  [0]{URL }%
\providecommand \Eprint [0]{\href }%
\providecommand \doibase [0]{https://doi.org/}%
\providecommand \selectlanguage [0]{\@gobble}%
\providecommand \bibinfo  [0]{\@secondoftwo}%
\providecommand \bibfield  [0]{\@secondoftwo}%
\providecommand \translation [1]{[#1]}%
\providecommand \BibitemOpen [0]{}%
\providecommand \bibitemStop [0]{}%
\providecommand \bibitemNoStop [0]{.\EOS\space}%
\providecommand \EOS [0]{\spacefactor3000\relax}%
\providecommand \BibitemShut  [1]{\csname bibitem#1\endcsname}%
\let\auto@bib@innerbib\@empty
%</preamble>
\bibitem [{\citenamefont {Mazziotti}(2012)}]{doi:10.1021/cr2000493}%
  \BibitemOpen
  \bibfield  {author} {\bibinfo {author} {\bibfnamefont {D.~A.}\ \bibnamefont
  {Mazziotti}},\ }\href {https://doi.org/10.1021/cr2000493} {\bibfield
  {journal} {\bibinfo  {journal} {Chemical Reviews}\ }\textbf {\bibinfo
  {volume} {112}},\ \bibinfo {pages} {244} (\bibinfo {year}
  {2012})}\BibitemShut {NoStop}%
\bibitem [{\citenamefont {Luzanov}\ and\ \citenamefont
  {Whyman}(1981)}]{https://doi.org/10.1002/qua.560200604}%
  \BibitemOpen
  \bibfield  {author} {\bibinfo {author} {\bibfnamefont {A.~V.}\ \bibnamefont
  {Luzanov}}\ and\ \bibinfo {author} {\bibfnamefont {G.~E.}\ \bibnamefont
  {Whyman}},\ }\href {https://doi.org/10.1002/qua.560200604} {\bibfield
  {journal} {\bibinfo  {journal} {International Journal of Quantum Chemistry}\
  }\textbf {\bibinfo {volume} {20}},\ \bibinfo {pages} {1179} (\bibinfo {year}
  {1981})}\BibitemShut {NoStop}%
\bibitem [{\citenamefont {Kutzelnigg}, \citenamefont {Shamasundar},\ and\
  \citenamefont {Mukherjee}(2010)}]{Kutzelnigg10022010}%
  \BibitemOpen
  \bibfield  {author} {\bibinfo {author} {\bibfnamefont {W.}~\bibnamefont
  {Kutzelnigg}}, \bibinfo {author} {\bibfnamefont {K.}~\bibnamefont
  {Shamasundar}},\ and\ \bibinfo {author} {\bibfnamefont {D.}~\bibnamefont
  {Mukherjee}},\ }\href {https://doi.org/10.1080/00268970903547926} {\bibfield
  {journal} {\bibinfo  {journal} {Molecular Physics}\ }\textbf {\bibinfo
  {volume} {108}},\ \bibinfo {pages} {433} (\bibinfo {year}
  {2010})}\BibitemShut {NoStop}%
\bibitem [{\citenamefont {Kutzelnigg}\ and\ \citenamefont
  {Mukherjee}(1999)}]{10.1063/1.478189}%
  \BibitemOpen
  \bibfield  {author} {\bibinfo {author} {\bibfnamefont {W.}~\bibnamefont
  {Kutzelnigg}}\ and\ \bibinfo {author} {\bibfnamefont {D.}~\bibnamefont
  {Mukherjee}},\ }\href {https://doi.org/10.1063/1.478189} {\bibfield
  {journal} {\bibinfo  {journal} {The Journal of Chemical Physics}\ }\textbf
  {\bibinfo {volume} {110}},\ \bibinfo {pages} {2800} (\bibinfo {year}
  {1999})}\BibitemShut {NoStop}%
\bibitem [{\citenamefont {B.}\ \emph {et~al.}(2026)\citenamefont {B.},
  \citenamefont {Rana}, \citenamefont {Shao}, \citenamefont {Pernal},\ and\
  \citenamefont {Pavanello}}]{b2026machinelearningtwoelectronreduced}%
  \BibitemOpen
  \bibfield  {author} {\bibinfo {author} {\bibfnamefont {J.~A.~M.}\
  \bibnamefont {B.}}, \bibinfo {author} {\bibfnamefont {B.}~\bibnamefont
  {Rana}}, \bibinfo {author} {\bibfnamefont {X.}~\bibnamefont {Shao}}, \bibinfo
  {author} {\bibfnamefont {K.}~\bibnamefont {Pernal}},\ and\ \bibinfo {author}
  {\bibfnamefont {M.}~\bibnamefont {Pavanello}},\ }\href
  {https://doi.org/10.48550/arxiv.2603.06882} {\enquote {\bibinfo {title}
  {Machine learning the two-electron reduced density matrix in molecules and
  condensed phases},}\ } (\bibinfo {year} {2026}),\ \Eprint
  {https://arxiv.org/abs/2603.06882} {arXiv:2603.06882 [physics.chem-ph]}
  \BibitemShut {NoStop}%
\bibitem [{\citenamefont {Raeber}\ and\ \citenamefont
  {Mazziotti}(2015)}]{PhysRevA.92.052502}%
  \BibitemOpen
  \bibfield  {author} {\bibinfo {author} {\bibfnamefont {A.}~\bibnamefont
  {Raeber}}\ and\ \bibinfo {author} {\bibfnamefont {D.~A.}\ \bibnamefont
  {Mazziotti}},\ }\href {https://doi.org/10.1103/PhysRevA.92.052502} {\bibfield
   {journal} {\bibinfo  {journal} {Phys. Rev. A}\ }\textbf {\bibinfo {volume}
  {92}},\ \bibinfo {pages} {052502} (\bibinfo {year} {2015})}\BibitemShut
  {NoStop}%
\bibitem [{\citenamefont {Shamasundar}(2009)}]{10.1063/1.3256237}%
  \BibitemOpen
  \bibfield  {author} {\bibinfo {author} {\bibfnamefont {K.~R.}\ \bibnamefont
  {Shamasundar}},\ }\href {https://doi.org/10.1063/1.3256237} {\bibfield
  {journal} {\bibinfo  {journal} {The Journal of Chemical Physics}\ }\textbf
  {\bibinfo {volume} {131}},\ \bibinfo {pages} {174109} (\bibinfo {year}
  {2009})}\BibitemShut {NoStop}%
\bibitem [{\citenamefont {Tel}\ \emph {et~al.}(2003)\citenamefont {Tel},
  \citenamefont {P\'erez-Romero}, \citenamefont {Casquero},\ and\ \citenamefont
  {Valdemoro}}]{PhysRevA.67.052504}%
  \BibitemOpen
  \bibfield  {author} {\bibinfo {author} {\bibfnamefont {L.~M.}\ \bibnamefont
  {Tel}}, \bibinfo {author} {\bibfnamefont {E.}~\bibnamefont {P\'erez-Romero}},
  \bibinfo {author} {\bibfnamefont {F.~J.}\ \bibnamefont {Casquero}},\ and\
  \bibinfo {author} {\bibfnamefont {C.}~\bibnamefont {Valdemoro}},\ }\href
  {https://doi.org/10.1103/PhysRevA.67.052504} {\bibfield  {journal} {\bibinfo
  {journal} {Phys. Rev. A}\ }\textbf {\bibinfo {volume} {67}},\ \bibinfo
  {pages} {052504} (\bibinfo {year} {2003})}\BibitemShut {NoStop}%
\bibitem [{\citenamefont {Schwerdtfeger}\ and\ \citenamefont
  {Mazziotti}(2012)}]{Mazziotti12-lowrank}%
  \BibitemOpen
  \bibfield  {author} {\bibinfo {author} {\bibfnamefont {C.~A.}\ \bibnamefont
  {Schwerdtfeger}}\ and\ \bibinfo {author} {\bibfnamefont {D.~A.}\ \bibnamefont
  {Mazziotti}},\ }\href {https://doi.org/10.1063/1.4770278} {\bibfield
  {journal} {\bibinfo  {journal} {The Journal of Chemical Physics}\ }\textbf
  {\bibinfo {volume} {137}},\ \bibinfo {pages} {244103} (\bibinfo {year}
  {2012})}\BibitemShut {NoStop}%
\bibitem [{\citenamefont {Kinoshita}, \citenamefont {Hino},\ and\ \citenamefont
  {Bartlett}(2003)}]{Kinoshita03-SVD-CCSD}%
  \BibitemOpen
  \bibfield  {author} {\bibinfo {author} {\bibfnamefont {T.}~\bibnamefont
  {Kinoshita}}, \bibinfo {author} {\bibfnamefont {O.}~\bibnamefont {Hino}},\
  and\ \bibinfo {author} {\bibfnamefont {R.~J.}\ \bibnamefont {Bartlett}},\
  }\href {https://doi.org/10.1063/1.1609442} {\bibfield  {journal} {\bibinfo
  {journal} {The Journal of Chemical Physics}\ }\textbf {\bibinfo {volume}
  {119}},\ \bibinfo {pages} {7756} (\bibinfo {year} {2003})}\BibitemShut
  {NoStop}%
\bibitem [{\citenamefont {Beran}\ and\ \citenamefont
  {Head-Gordon}(2004)}]{HeadGordon04-SVD-T2}%
  \BibitemOpen
  \bibfield  {author} {\bibinfo {author} {\bibfnamefont {G.~J.~O.}\
  \bibnamefont {Beran}}\ and\ \bibinfo {author} {\bibfnamefont
  {M.}~\bibnamefont {Head-Gordon}},\ }\href {https://doi.org/10.1063/1.1756860}
  {\bibfield  {journal} {\bibinfo  {journal} {The Journal of Chemical Physics}\
  }\textbf {\bibinfo {volume} {121}},\ \bibinfo {pages} {78} (\bibinfo {year}
  {2004})}\BibitemShut {NoStop}%
\bibitem [{\citenamefont {Bell}, \citenamefont {Lambrecht},\ and\ \citenamefont
  {Head-Gordon}(2010)}]{Bell10-HOSVD-MP2}%
  \BibitemOpen
  \bibfield  {author} {\bibinfo {author} {\bibfnamefont {F.}~\bibnamefont
  {Bell}}, \bibinfo {author} {\bibfnamefont {D.}~\bibnamefont {Lambrecht}},\
  and\ \bibinfo {author} {\bibfnamefont {M.}~\bibnamefont {Head-Gordon}},\
  }\href {https://doi.org/10.1080/00268976.2010.523713} {\bibfield  {journal}
  {\bibinfo  {journal} {Molecular Physics}\ }\textbf {\bibinfo {volume}
  {108}},\ \bibinfo {pages} {2759} (\bibinfo {year} {2010})}\BibitemShut
  {NoStop}%
\bibitem [{\citenamefont {Benedikt}\ \emph {et~al.}(2011)\citenamefont
  {Benedikt}, \citenamefont {Auer}, \citenamefont {Espig},\ and\ \citenamefont
  {Hackbusch}}]{Benedikt11-CP-decomp-MP2}%
  \BibitemOpen
  \bibfield  {author} {\bibinfo {author} {\bibfnamefont {U.}~\bibnamefont
  {Benedikt}}, \bibinfo {author} {\bibfnamefont {A.~A.}\ \bibnamefont {Auer}},
  \bibinfo {author} {\bibfnamefont {M.}~\bibnamefont {Espig}},\ and\ \bibinfo
  {author} {\bibfnamefont {W.}~\bibnamefont {Hackbusch}},\ }\href
  {https://doi.org/10.1063/1.3514201} {\bibfield  {journal} {\bibinfo
  {journal} {The Journal of Chemical Physics}\ }\textbf {\bibinfo {volume}
  {134}},\ \bibinfo {pages} {054118} (\bibinfo {year} {2011})}\BibitemShut
  {NoStop}%
\bibitem [{\citenamefont {Yang}\ \emph {et~al.}(2012)\citenamefont {Yang},
  \citenamefont {Chan}, \citenamefont {Manby}, \citenamefont {Schütz},\ and\
  \citenamefont {Werner}}]{Yang12-virtualorbopt-CCSD}%
  \BibitemOpen
  \bibfield  {author} {\bibinfo {author} {\bibfnamefont {J.}~\bibnamefont
  {Yang}}, \bibinfo {author} {\bibfnamefont {G.~K.-L.}\ \bibnamefont {Chan}},
  \bibinfo {author} {\bibfnamefont {F.~R.}\ \bibnamefont {Manby}}, \bibinfo
  {author} {\bibfnamefont {M.}~\bibnamefont {Schütz}},\ and\ \bibinfo {author}
  {\bibfnamefont {H.-J.}\ \bibnamefont {Werner}},\ }\href
  {https://doi.org/10.1063/1.3696963} {\bibfield  {journal} {\bibinfo
  {journal} {The Journal of Chemical Physics}\ }\textbf {\bibinfo {volume}
  {136}},\ \bibinfo {pages} {144105} (\bibinfo {year} {2012})}\BibitemShut
  {NoStop}%
\bibitem [{\citenamefont {Scott}\ and\ \citenamefont
  {Booth}(2021)}]{PhysRevB.104.245114}%
  \BibitemOpen
  \bibfield  {author} {\bibinfo {author} {\bibfnamefont {C.~J.~C.}\
  \bibnamefont {Scott}}\ and\ \bibinfo {author} {\bibfnamefont {G.~H.}\
  \bibnamefont {Booth}},\ }\href {https://doi.org/10.1103/PhysRevB.104.245114}
  {\bibfield  {journal} {\bibinfo  {journal} {Phys. Rev. B}\ }\textbf {\bibinfo
  {volume} {104}},\ \bibinfo {pages} {245114} (\bibinfo {year}
  {2021})}\BibitemShut {NoStop}%
\bibitem [{\citenamefont {Parrish}\ \emph {et~al.}(2019)\citenamefont
  {Parrish}, \citenamefont {Zhao}, \citenamefont {Hohenstein},\ and\
  \citenamefont {Martínez}}]{10.1063/1.5092505}%
  \BibitemOpen
  \bibfield  {author} {\bibinfo {author} {\bibfnamefont {R.~M.}\ \bibnamefont
  {Parrish}}, \bibinfo {author} {\bibfnamefont {Y.}~\bibnamefont {Zhao}},
  \bibinfo {author} {\bibfnamefont {E.~G.}\ \bibnamefont {Hohenstein}},\ and\
  \bibinfo {author} {\bibfnamefont {T.~J.}\ \bibnamefont {Martínez}},\ }\href
  {https://doi.org/10.1063/1.5092505} {\bibfield  {journal} {\bibinfo
  {journal} {The Journal of Chemical Physics}\ }\textbf {\bibinfo {volume}
  {150}},\ \bibinfo {pages} {164118} (\bibinfo {year} {2019})}\BibitemShut
  {NoStop}%
\bibitem [{\citenamefont {Hohenstein}\ \emph {et~al.}(2012)\citenamefont
  {Hohenstein}, \citenamefont {Parrish}, \citenamefont {Sherrill},\ and\
  \citenamefont {Martínez}}]{Martinez12-THC-T2}%
  \BibitemOpen
  \bibfield  {author} {\bibinfo {author} {\bibfnamefont {E.~G.}\ \bibnamefont
  {Hohenstein}}, \bibinfo {author} {\bibfnamefont {R.~M.}\ \bibnamefont
  {Parrish}}, \bibinfo {author} {\bibfnamefont {C.~D.}\ \bibnamefont
  {Sherrill}},\ and\ \bibinfo {author} {\bibfnamefont {T.~J.}\ \bibnamefont
  {Martínez}},\ }\href {https://doi.org/10.1063/1.4768241} {\bibfield
  {journal} {\bibinfo  {journal} {The Journal of Chemical Physics}\ }\textbf
  {\bibinfo {volume} {137}},\ \bibinfo {pages} {221101} (\bibinfo {year}
  {2012})}\BibitemShut {NoStop}%
\bibitem [{\citenamefont {Nusspickel}\ and\ \citenamefont
  {Booth}(2020)}]{PhysRevB.102.165107}%
  \BibitemOpen
  \bibfield  {author} {\bibinfo {author} {\bibfnamefont {M.}~\bibnamefont
  {Nusspickel}}\ and\ \bibinfo {author} {\bibfnamefont {G.~H.}\ \bibnamefont
  {Booth}},\ }\href {https://doi.org/10.1103/PhysRevB.102.165107} {\bibfield
  {journal} {\bibinfo  {journal} {Phys. Rev. B}\ }\textbf {\bibinfo {volume}
  {102}},\ \bibinfo {pages} {165107} (\bibinfo {year} {2020})}\BibitemShut
  {NoStop}%
\bibitem [{\citenamefont {Mazziotti}(2006)}]{Mazziotti06-v2RDM}%
  \BibitemOpen
  \bibfield  {author} {\bibinfo {author} {\bibfnamefont {D.~A.}\ \bibnamefont
  {Mazziotti}},\ }\href {https://doi.org/10.1021/ar050029d} {\bibfield
  {journal} {\bibinfo  {journal} {Accounts of Chemical Research}\ }\textbf
  {\bibinfo {volume} {39}},\ \bibinfo {pages} {207} (\bibinfo {year}
  {2006})}\BibitemShut {NoStop}%
\bibitem [{\citenamefont {DePrince}\ and\ \citenamefont
  {Mazziotti}(2007)}]{Mazziotti07-parametric2RDM}%
  \BibitemOpen
  \bibfield  {author} {\bibinfo {author} {\bibfnamefont {A.~E.}\ \bibnamefont
  {DePrince}}\ and\ \bibinfo {author} {\bibfnamefont {D.~A.}\ \bibnamefont
  {Mazziotti}},\ }\href {https://doi.org/10.1103/PhysRevA.76.042501} {\bibfield
   {journal} {\bibinfo  {journal} {Phys. Rev. A}\ }\textbf {\bibinfo {volume}
  {76}},\ \bibinfo {pages} {042501} (\bibinfo {year} {2007})}\BibitemShut
  {NoStop}%
\bibitem [{\citenamefont {Eugene
  DePrince~III}(2024)}]{https://doi.org/10.1002/wcms.1702}%
  \BibitemOpen
  \bibfield  {author} {\bibinfo {author} {\bibfnamefont {A.}~\bibnamefont
  {Eugene DePrince~III}},\ }\href {https://doi.org/10.1002/wcms.1702}
  {\bibfield  {journal} {\bibinfo  {journal} {WIREs Computational Molecular
  Science}\ }\textbf {\bibinfo {volume} {14}},\ \bibinfo {pages} {e1702}
  (\bibinfo {year} {2024})}\BibitemShut {NoStop}%
\bibitem [{\citenamefont {Gidofalvi}\ and\ \citenamefont
  {Mazziotti}(2007)}]{Mazziotti07-lowrank-v2RDM}%
  \BibitemOpen
  \bibfield  {author} {\bibinfo {author} {\bibfnamefont {G.}~\bibnamefont
  {Gidofalvi}}\ and\ \bibinfo {author} {\bibfnamefont {D.~A.}\ \bibnamefont
  {Mazziotti}},\ }\href {https://doi.org/10.1063/1.2817602} {\bibfield
  {journal} {\bibinfo  {journal} {The Journal of Chemical Physics}\ }\textbf
  {\bibinfo {volume} {127}},\ \bibinfo {pages} {244105} (\bibinfo {year}
  {2007})}\BibitemShut {NoStop}%
\bibitem [{\citenamefont {Hoy}, \citenamefont {Shenvi},\ and\ \citenamefont
  {Mazziotti}(2013)}]{Mazziotti13-comparison_lowrank}%
  \BibitemOpen
  \bibfield  {author} {\bibinfo {author} {\bibfnamefont {E.~P.}\ \bibnamefont
  {Hoy}}, \bibinfo {author} {\bibfnamefont {N.}~\bibnamefont {Shenvi}},\ and\
  \bibinfo {author} {\bibfnamefont {D.~A.}\ \bibnamefont {Mazziotti}},\ }\href
  {https://doi.org/10.1063/1.4813495} {\bibfield  {journal} {\bibinfo
  {journal} {The Journal of Chemical Physics}\ }\textbf {\bibinfo {volume}
  {139}},\ \bibinfo {pages} {034105} (\bibinfo {year} {2013})}\BibitemShut
  {NoStop}%
\bibitem [{\citenamefont {Massaccesi}\ \emph {et~al.}(2026)\citenamefont
  {Massaccesi}, \citenamefont {O{\~n}a}, \citenamefont {Lain}, \citenamefont
  {Torre}, \citenamefont {Peralta}, \citenamefont {Alcoba},\ and\ \citenamefont
  {Scuseria}}]{doi:10.1021/acs.jpclett.6c00296}%
  \BibitemOpen
  \bibfield  {author} {\bibinfo {author} {\bibfnamefont {G.~E.}\ \bibnamefont
  {Massaccesi}}, \bibinfo {author} {\bibfnamefont {O.~B.}\ \bibnamefont
  {O{\~n}a}}, \bibinfo {author} {\bibfnamefont {L.}~\bibnamefont {Lain}},
  \bibinfo {author} {\bibfnamefont {A.}~\bibnamefont {Torre}}, \bibinfo
  {author} {\bibfnamefont {J.~E.}\ \bibnamefont {Peralta}}, \bibinfo {author}
  {\bibfnamefont {D.~R.}\ \bibnamefont {Alcoba}},\ and\ \bibinfo {author}
  {\bibfnamefont {G.~E.}\ \bibnamefont {Scuseria}},\ }\href
  {https://doi.org/10.1021/acs.jpclett.6c00296} {\bibfield  {journal} {\bibinfo
   {journal} {The Journal of Physical Chemistry Letters}\ }\textbf {\bibinfo
  {volume} {17}},\ \bibinfo {pages} {3430} (\bibinfo {year}
  {2026})}\BibitemShut {NoStop}%
\bibitem [{\citenamefont {Peng}, \citenamefont {Zhang},\ and\ \citenamefont
  {Chan}(2023)}]{Chan23-RDM-completion}%
  \BibitemOpen
  \bibfield  {author} {\bibinfo {author} {\bibfnamefont {L.}~\bibnamefont
  {Peng}}, \bibinfo {author} {\bibfnamefont {X.}~\bibnamefont {Zhang}},\ and\
  \bibinfo {author} {\bibfnamefont {G.~K.-L.}\ \bibnamefont {Chan}},\ }\href
  {https://doi.org/10.1021/acs.jctc.3c00851} {\bibfield  {journal} {\bibinfo
  {journal} {Journal of Chemical Theory and Computation}\ }\textbf {\bibinfo
  {volume} {19}},\ \bibinfo {pages} {9151} (\bibinfo {year}
  {2023})}\BibitemShut {NoStop}%
\bibitem [{\citenamefont {Anselmetti}\ \emph {et~al.}(2025)\citenamefont
  {Anselmetti}, \citenamefont {Degroote}, \citenamefont {Moll}, \citenamefont
  {Santagati},\ and\ \citenamefont
  {Streif}}]{anselmetti2025classicaloptimisationreduceddensity}%
  \BibitemOpen
  \bibfield  {author} {\bibinfo {author} {\bibfnamefont {G.-L.~R.}\
  \bibnamefont {Anselmetti}}, \bibinfo {author} {\bibfnamefont
  {M.}~\bibnamefont {Degroote}}, \bibinfo {author} {\bibfnamefont
  {N.}~\bibnamefont {Moll}}, \bibinfo {author} {\bibfnamefont {R.}~\bibnamefont
  {Santagati}},\ and\ \bibinfo {author} {\bibfnamefont {M.}~\bibnamefont
  {Streif}},\ }\href {https://doi.org/10.48550/arxiv.2411.18430} {\enquote
  {\bibinfo {title} {Classical optimisation of reduced density matrix
  estimations with classical shadows using n-representability conditions under
  shot noise considerations},}\ } (\bibinfo {year} {2025}),\ \Eprint
  {https://arxiv.org/abs/2411.18430} {arXiv:2411.18430 [quant-ph]} \BibitemShut
  {NoStop}%
\bibitem [{\citenamefont {Bartlett}\ and\ \citenamefont
  {Musia\l{}}(2007)}]{RevModPhys.79.291}%
  \BibitemOpen
  \bibfield  {author} {\bibinfo {author} {\bibfnamefont {R.~J.}\ \bibnamefont
  {Bartlett}}\ and\ \bibinfo {author} {\bibfnamefont {M.}~\bibnamefont
  {Musia\l{}}},\ }\href {https://doi.org/10.1103/RevModPhys.79.291} {\bibfield
  {journal} {\bibinfo  {journal} {Rev. Mod. Phys.}\ }\textbf {\bibinfo {volume}
  {79}},\ \bibinfo {pages} {291} (\bibinfo {year} {2007})}\BibitemShut
  {NoStop}%
\bibitem [{\citenamefont {Zgid}\ and\ \citenamefont
  {Nooijen}(2008)}]{10.1063/1.2883980}%
  \BibitemOpen
  \bibfield  {author} {\bibinfo {author} {\bibfnamefont {D.}~\bibnamefont
  {Zgid}}\ and\ \bibinfo {author} {\bibfnamefont {M.}~\bibnamefont {Nooijen}},\
  }\href {https://doi.org/10.1063/1.2883980} {\bibfield  {journal} {\bibinfo
  {journal} {The Journal of Chemical Physics}\ }\textbf {\bibinfo {volume}
  {128}},\ \bibinfo {pages} {144115} (\bibinfo {year} {2008})}\BibitemShut
  {NoStop}%
\bibitem [{\citenamefont {Duguet}\ \emph {et~al.}(2024)\citenamefont {Duguet},
  \citenamefont {Ekstr\"om}, \citenamefont {Furnstahl}, \citenamefont
  {K\"onig},\ and\ \citenamefont {Lee}}]{RevModPhys.96.031002}%
  \BibitemOpen
  \bibfield  {author} {\bibinfo {author} {\bibfnamefont {T.}~\bibnamefont
  {Duguet}}, \bibinfo {author} {\bibfnamefont {A.}~\bibnamefont {Ekstr\"om}},
  \bibinfo {author} {\bibfnamefont {R.~J.}\ \bibnamefont {Furnstahl}}, \bibinfo
  {author} {\bibfnamefont {S.}~\bibnamefont {K\"onig}},\ and\ \bibinfo {author}
  {\bibfnamefont {D.}~\bibnamefont {Lee}},\ }\href
  {https://doi.org/10.1103/RevModPhys.96.031002} {\bibfield  {journal}
  {\bibinfo  {journal} {Rev. Mod. Phys.}\ }\textbf {\bibinfo {volume} {96}},\
  \bibinfo {pages} {031002} (\bibinfo {year} {2024})}\BibitemShut {NoStop}%
\bibitem [{\citenamefont {Mejuto-Zaera}\ and\ \citenamefont
  {Kemper}(2023)}]{mejutozaera2023-evcont}%
  \BibitemOpen
  \bibfield  {author} {\bibinfo {author} {\bibfnamefont {C.}~\bibnamefont
  {Mejuto-Zaera}}\ and\ \bibinfo {author} {\bibfnamefont {A.~F.}\ \bibnamefont
  {Kemper}},\ }\href {https://doi.org/10.1088/2516-1075/ad018f} {\bibfield
  {journal} {\bibinfo  {journal} {Electron. Struct.}\ }\textbf {\bibinfo
  {volume} {5}},\ \bibinfo {pages} {045007} (\bibinfo {year}
  {2023})}\BibitemShut {NoStop}%
\bibitem [{\citenamefont {Rath}\ and\ \citenamefont
  {Booth}(2025)}]{Rath25-evcont}%
  \BibitemOpen
  \bibfield  {author} {\bibinfo {author} {\bibfnamefont {Y.}~\bibnamefont
  {Rath}}\ and\ \bibinfo {author} {\bibfnamefont {G.~H.}\ \bibnamefont
  {Booth}},\ }\href {https://doi.org/10.1038/s41467-025-57134-9} {\bibfield
  {journal} {\bibinfo  {journal} {Nature Communications}\ }\textbf {\bibinfo
  {volume} {16}},\ \bibinfo {pages} {2005} (\bibinfo {year}
  {2025})}\BibitemShut {NoStop}%
\bibitem [{\citenamefont {Atalar}\ \emph {et~al.}(2024)\citenamefont {Atalar},
  \citenamefont {Rath}, \citenamefont {Crespo-Otero},\ and\ \citenamefont
  {Booth}}]{Atalar24-eigconNAMD-faraday}%
  \BibitemOpen
  \bibfield  {author} {\bibinfo {author} {\bibfnamefont {K.}~\bibnamefont
  {Atalar}}, \bibinfo {author} {\bibfnamefont {Y.}~\bibnamefont {Rath}},
  \bibinfo {author} {\bibfnamefont {R.}~\bibnamefont {Crespo-Otero}},\ and\
  \bibinfo {author} {\bibfnamefont {G.~H.}\ \bibnamefont {Booth}},\ }\href
  {https://doi.org/10.1039/D4FD00062E} {\bibfield  {journal} {\bibinfo
  {journal} {Faraday Discuss.}\ }\textbf {\bibinfo {volume} {254}},\ \bibinfo
  {pages} {542} (\bibinfo {year} {2024})}\BibitemShut {NoStop}%
\bibitem [{\citenamefont {Burton}(2021)}]{10.1063/5.0045442}%
  \BibitemOpen
  \bibfield  {author} {\bibinfo {author} {\bibfnamefont {H.~G.~A.}\
  \bibnamefont {Burton}},\ }\href {https://doi.org/10.1063/5.0045442}
  {\bibfield  {journal} {\bibinfo  {journal} {The Journal of Chemical Physics}\
  }\textbf {\bibinfo {volume} {154}},\ \bibinfo {pages} {144109} (\bibinfo
  {year} {2021})}\BibitemShut {NoStop}%
\bibitem [{\citenamefont {Burton}(2022)}]{10.1063/5.0122094}%
  \BibitemOpen
  \bibfield  {author} {\bibinfo {author} {\bibfnamefont {H.~G.~A.}\
  \bibnamefont {Burton}},\ }\href {https://doi.org/10.1063/5.0122094}
  {\bibfield  {journal} {\bibinfo  {journal} {The Journal of Chemical Physics}\
  }\textbf {\bibinfo {volume} {157}},\ \bibinfo {pages} {204109} (\bibinfo
  {year} {2022})}\BibitemShut {NoStop}%
\bibitem [{\citenamefont {Harris}(2002)}]{https://doi.org/10.1002/qua.997}%
  \BibitemOpen
  \bibfield  {author} {\bibinfo {author} {\bibfnamefont {F.~E.}\ \bibnamefont
  {Harris}},\ }\href {https://doi.org/10.1002/qua.997} {\bibfield  {journal}
  {\bibinfo  {journal} {International Journal of Quantum Chemistry}\ }\textbf
  {\bibinfo {volume} {90}},\ \bibinfo {pages} {105} (\bibinfo {year}
  {2002})}\BibitemShut {NoStop}%
\bibitem [{\citenamefont {McWeeny}\ and\ \citenamefont
  {Kutzelnigg}(1968)}]{https://doi.org/10.1002/qua.560020203}%
  \BibitemOpen
  \bibfield  {author} {\bibinfo {author} {\bibfnamefont {R.}~\bibnamefont
  {McWeeny}}\ and\ \bibinfo {author} {\bibfnamefont {W.}~\bibnamefont
  {Kutzelnigg}},\ }\href {https://doi.org/10.1002/qua.560020203} {\bibfield
  {journal} {\bibinfo  {journal} {International Journal of Quantum Chemistry}\
  }\textbf {\bibinfo {volume} {2}},\ \bibinfo {pages} {187} (\bibinfo {year}
  {1968})}\BibitemShut {NoStop}%
\bibitem [{\citenamefont {Hammond}\ and\ \citenamefont
  {Mazziotti}(2005)}]{PhysRevA.71.062503}%
  \BibitemOpen
  \bibfield  {author} {\bibinfo {author} {\bibfnamefont {J.~R.}\ \bibnamefont
  {Hammond}}\ and\ \bibinfo {author} {\bibfnamefont {D.~A.}\ \bibnamefont
  {Mazziotti}},\ }\href {https://doi.org/10.1103/PhysRevA.71.062503} {\bibfield
   {journal} {\bibinfo  {journal} {Phys. Rev. A}\ }\textbf {\bibinfo {volume}
  {71}},\ \bibinfo {pages} {062503} (\bibinfo {year} {2005})}\BibitemShut
  {NoStop}%
\bibitem [{\citenamefont {Barnett}\ and\ \citenamefont
  {Platas}(1968)}]{10.1063/1.1669767}%
  \BibitemOpen
  \bibfield  {author} {\bibinfo {author} {\bibfnamefont {G.~P.}\ \bibnamefont
  {Barnett}}\ and\ \bibinfo {author} {\bibfnamefont {O.~R.}\ \bibnamefont
  {Platas}},\ }\href {https://doi.org/10.1063/1.1669767} {\bibfield  {journal}
  {\bibinfo  {journal} {The Journal of Chemical Physics}\ }\textbf {\bibinfo
  {volume} {48}},\ \bibinfo {pages} {4265} (\bibinfo {year}
  {1968})}\BibitemShut {NoStop}%
\bibitem [{\citenamefont {Sun}\ \emph {et~al.}(2018)\citenamefont {Sun},
  \citenamefont {Berkelbach}, \citenamefont {Blunt}, \citenamefont {Booth},
  \citenamefont {Guo}, \citenamefont {Li}, \citenamefont {Liu}, \citenamefont
  {McClain}, \citenamefont {Sayfutyarova}, \citenamefont {Sharma},
  \citenamefont {Wouters},\ and\ \citenamefont {Chan}}]{pyscf2018}%
  \BibitemOpen
  \bibfield  {author} {\bibinfo {author} {\bibfnamefont {Q.}~\bibnamefont
  {Sun}}, \bibinfo {author} {\bibfnamefont {T.~C.}\ \bibnamefont {Berkelbach}},
  \bibinfo {author} {\bibfnamefont {N.~S.}\ \bibnamefont {Blunt}}, \bibinfo
  {author} {\bibfnamefont {G.~H.}\ \bibnamefont {Booth}}, \bibinfo {author}
  {\bibfnamefont {S.}~\bibnamefont {Guo}}, \bibinfo {author} {\bibfnamefont
  {Z.}~\bibnamefont {Li}}, \bibinfo {author} {\bibfnamefont {J.}~\bibnamefont
  {Liu}}, \bibinfo {author} {\bibfnamefont {J.~D.}\ \bibnamefont {McClain}},
  \bibinfo {author} {\bibfnamefont {E.~R.}\ \bibnamefont {Sayfutyarova}},
  \bibinfo {author} {\bibfnamefont {S.}~\bibnamefont {Sharma}}, \bibinfo
  {author} {\bibfnamefont {S.}~\bibnamefont {Wouters}},\ and\ \bibinfo {author}
  {\bibfnamefont {G.~K.-L.}\ \bibnamefont {Chan}},\ }\href
  {https://doi.org/10.1002/wcms.1340} {\bibfield  {journal} {\bibinfo
  {journal} {WIREs Comput. Mol. Sci.}\ }\textbf {\bibinfo {volume} {8}},\
  \bibinfo {pages} {e1340} (\bibinfo {year} {2018})}\BibitemShut {NoStop}%
\bibitem [{\citenamefont {Sun}\ \emph {et~al.}(2020)\citenamefont {Sun},
  \citenamefont {Zhang}, \citenamefont {Banerjee}, \citenamefont {Bao},
  \citenamefont {Barbry}, \citenamefont {Blunt}, \citenamefont {Bogdanov},
  \citenamefont {Booth}, \citenamefont {Chen}, \citenamefont {Cui},
  \citenamefont {Eriksen}, \citenamefont {Gao}, \citenamefont {Guo},
  \citenamefont {Hermann}, \citenamefont {Hermes}, \citenamefont {Koh},
  \citenamefont {Koval}, \citenamefont {Lehtola}, \citenamefont {Li},
  \citenamefont {Liu}, \citenamefont {Mardirossian}, \citenamefont {McClain},
  \citenamefont {Motta}, \citenamefont {Mussard}, \citenamefont {Pham},
  \citenamefont {Pulkin}, \citenamefont {Purwanto}, \citenamefont {Robinson},
  \citenamefont {Ronca}, \citenamefont {Sayfutyarova}, \citenamefont
  {Scheurer}, \citenamefont {Schurkus}, \citenamefont {Smith}, \citenamefont
  {Sun}, \citenamefont {Sun}, \citenamefont {Upadhyay}, \citenamefont {Wagner},
  \citenamefont {Wang}, \citenamefont {White}, \citenamefont {Whitfield},
  \citenamefont {Williamson}, \citenamefont {Wouters}, \citenamefont {Yang},
  \citenamefont {Yu}, \citenamefont {Zhu}, \citenamefont {Berkelbach},
  \citenamefont {Sharma}, \citenamefont {Sokolov},\ and\ \citenamefont
  {Chan}}]{pyscf2020}%
  \BibitemOpen
  \bibfield  {author} {\bibinfo {author} {\bibfnamefont {Q.}~\bibnamefont
  {Sun}}, \bibinfo {author} {\bibfnamefont {X.}~\bibnamefont {Zhang}}, \bibinfo
  {author} {\bibfnamefont {S.}~\bibnamefont {Banerjee}}, \bibinfo {author}
  {\bibfnamefont {P.}~\bibnamefont {Bao}}, \bibinfo {author} {\bibfnamefont
  {M.}~\bibnamefont {Barbry}}, \bibinfo {author} {\bibfnamefont {N.~S.}\
  \bibnamefont {Blunt}}, \bibinfo {author} {\bibfnamefont {N.~A.}\ \bibnamefont
  {Bogdanov}}, \bibinfo {author} {\bibfnamefont {G.~H.}\ \bibnamefont {Booth}},
  \bibinfo {author} {\bibfnamefont {J.}~\bibnamefont {Chen}}, \bibinfo {author}
  {\bibfnamefont {Z.-H.}\ \bibnamefont {Cui}}, \bibinfo {author} {\bibfnamefont
  {J.~J.}\ \bibnamefont {Eriksen}}, \bibinfo {author} {\bibfnamefont
  {Y.}~\bibnamefont {Gao}}, \bibinfo {author} {\bibfnamefont {S.}~\bibnamefont
  {Guo}}, \bibinfo {author} {\bibfnamefont {J.}~\bibnamefont {Hermann}},
  \bibinfo {author} {\bibfnamefont {M.~R.}\ \bibnamefont {Hermes}}, \bibinfo
  {author} {\bibfnamefont {K.}~\bibnamefont {Koh}}, \bibinfo {author}
  {\bibfnamefont {P.}~\bibnamefont {Koval}}, \bibinfo {author} {\bibfnamefont
  {S.}~\bibnamefont {Lehtola}}, \bibinfo {author} {\bibfnamefont
  {Z.}~\bibnamefont {Li}}, \bibinfo {author} {\bibfnamefont {J.}~\bibnamefont
  {Liu}}, \bibinfo {author} {\bibfnamefont {N.}~\bibnamefont {Mardirossian}},
  \bibinfo {author} {\bibfnamefont {J.~D.}\ \bibnamefont {McClain}}, \bibinfo
  {author} {\bibfnamefont {M.}~\bibnamefont {Motta}}, \bibinfo {author}
  {\bibfnamefont {B.}~\bibnamefont {Mussard}}, \bibinfo {author} {\bibfnamefont
  {H.~Q.}\ \bibnamefont {Pham}}, \bibinfo {author} {\bibfnamefont
  {A.}~\bibnamefont {Pulkin}}, \bibinfo {author} {\bibfnamefont
  {W.}~\bibnamefont {Purwanto}}, \bibinfo {author} {\bibfnamefont {P.~J.}\
  \bibnamefont {Robinson}}, \bibinfo {author} {\bibfnamefont {E.}~\bibnamefont
  {Ronca}}, \bibinfo {author} {\bibfnamefont {E.~R.}\ \bibnamefont
  {Sayfutyarova}}, \bibinfo {author} {\bibfnamefont {M.}~\bibnamefont
  {Scheurer}}, \bibinfo {author} {\bibfnamefont {H.~F.}\ \bibnamefont
  {Schurkus}}, \bibinfo {author} {\bibfnamefont {J.~E.~T.}\ \bibnamefont
  {Smith}}, \bibinfo {author} {\bibfnamefont {C.}~\bibnamefont {Sun}}, \bibinfo
  {author} {\bibfnamefont {S.-N.}\ \bibnamefont {Sun}}, \bibinfo {author}
  {\bibfnamefont {S.}~\bibnamefont {Upadhyay}}, \bibinfo {author}
  {\bibfnamefont {L.~K.}\ \bibnamefont {Wagner}}, \bibinfo {author}
  {\bibfnamefont {X.}~\bibnamefont {Wang}}, \bibinfo {author} {\bibfnamefont
  {A.}~\bibnamefont {White}}, \bibinfo {author} {\bibfnamefont {J.~D.}\
  \bibnamefont {Whitfield}}, \bibinfo {author} {\bibfnamefont {M.~J.}\
  \bibnamefont {Williamson}}, \bibinfo {author} {\bibfnamefont
  {S.}~\bibnamefont {Wouters}}, \bibinfo {author} {\bibfnamefont
  {J.}~\bibnamefont {Yang}}, \bibinfo {author} {\bibfnamefont {J.~M.}\
  \bibnamefont {Yu}}, \bibinfo {author} {\bibfnamefont {T.}~\bibnamefont
  {Zhu}}, \bibinfo {author} {\bibfnamefont {T.~C.}\ \bibnamefont {Berkelbach}},
  \bibinfo {author} {\bibfnamefont {S.}~\bibnamefont {Sharma}}, \bibinfo
  {author} {\bibfnamefont {A.~Y.}\ \bibnamefont {Sokolov}},\ and\ \bibinfo
  {author} {\bibfnamefont {G.~K.-L.}\ \bibnamefont {Chan}},\ }\href
  {https://doi.org/10.1063/5.0006074} {\bibfield  {journal} {\bibinfo
  {journal} {J. Chem. Phys.}\ }\textbf {\bibinfo {volume} {153}},\ \bibinfo
  {pages} {024109} (\bibinfo {year} {2020})}\BibitemShut {NoStop}%
\bibitem [{\citenamefont {Luzanov}(2012)}]{https://doi.org/10.1002/qua.24101}%
  \BibitemOpen
  \bibfield  {author} {\bibinfo {author} {\bibfnamefont {A.~V.}\ \bibnamefont
  {Luzanov}},\ }\href {https://doi.org/10.1002/qua.24101} {\bibfield  {journal}
  {\bibinfo  {journal} {International Journal of Quantum Chemistry}\ }\textbf
  {\bibinfo {volume} {112}},\ \bibinfo {pages} {2915} (\bibinfo {year}
  {2012})}\BibitemShut {NoStop}%
\bibitem [{\citenamefont {Bistoni}\ \emph {et~al.}(2024)\citenamefont
  {Bistoni}, \citenamefont {Altun}, \citenamefont {Wang},\ and\ \citenamefont
  {Neese}}]{doi:10.1021/acs.accounts.4c00085}%
  \BibitemOpen
  \bibfield  {author} {\bibinfo {author} {\bibfnamefont {G.}~\bibnamefont
  {Bistoni}}, \bibinfo {author} {\bibfnamefont {A.}~\bibnamefont {Altun}},
  \bibinfo {author} {\bibfnamefont {Z.}~\bibnamefont {Wang}},\ and\ \bibinfo
  {author} {\bibfnamefont {F.}~\bibnamefont {Neese}},\ }\href
  {https://doi.org/10.1021/acs.accounts.4c00085} {\bibfield  {journal}
  {\bibinfo  {journal} {Accounts of Chemical Research}\ }\textbf {\bibinfo
  {volume} {57}},\ \bibinfo {pages} {1411} (\bibinfo {year} {2024})},\ \bibinfo
  {note} {pMID: 38602396}\BibitemShut {NoStop}%
\bibitem [{\citenamefont {Cand{\`e}s}\ and\ \citenamefont
  {Recht}(2009)}]{Candes2009}%
  \BibitemOpen
  \bibfield  {author} {\bibinfo {author} {\bibfnamefont {E.~J.}\ \bibnamefont
  {Cand{\`e}s}}\ and\ \bibinfo {author} {\bibfnamefont {B.}~\bibnamefont
  {Recht}},\ }\href {https://doi.org/10.1007/s10208-009-9045-5} {\bibfield
  {journal} {\bibinfo  {journal} {Foundations of Computational Mathematics}\
  }\textbf {\bibinfo {volume} {9}},\ \bibinfo {pages} {717} (\bibinfo {year}
  {2009})}\BibitemShut {NoStop}%
\bibitem [{\citenamefont {Wang}\ and\ \citenamefont
  {Song}(2016)}]{10.1063/1.4952956}%
  \BibitemOpen
  \bibfield  {author} {\bibinfo {author} {\bibfnamefont {L.-P.}\ \bibnamefont
  {Wang}}\ and\ \bibinfo {author} {\bibfnamefont {C.}~\bibnamefont {Song}},\
  }\href {https://doi.org/10.1063/1.4952956} {\bibfield  {journal} {\bibinfo
  {journal} {The Journal of Chemical Physics}\ }\textbf {\bibinfo {volume}
  {144}},\ \bibinfo {pages} {214108} (\bibinfo {year} {2016})}\BibitemShut
  {NoStop}%
\bibitem [{\citenamefont {Reine}\ \emph {et~al.}(2008)\citenamefont {Reine},
  \citenamefont {Tellgren}, \citenamefont {Krapp}, \citenamefont {Kjærgaard},
  \citenamefont {Helgaker}, \citenamefont {Jansik}, \citenamefont {Høst},\
  and\ \citenamefont {Salek}}]{10.1063/1.2956507}%
  \BibitemOpen
  \bibfield  {author} {\bibinfo {author} {\bibfnamefont {S.}~\bibnamefont
  {Reine}}, \bibinfo {author} {\bibfnamefont {E.}~\bibnamefont {Tellgren}},
  \bibinfo {author} {\bibfnamefont {A.}~\bibnamefont {Krapp}}, \bibinfo
  {author} {\bibfnamefont {T.}~\bibnamefont {Kjærgaard}}, \bibinfo {author}
  {\bibfnamefont {T.}~\bibnamefont {Helgaker}}, \bibinfo {author}
  {\bibfnamefont {B.}~\bibnamefont {Jansik}}, \bibinfo {author} {\bibfnamefont
  {S.}~\bibnamefont {Høst}},\ and\ \bibinfo {author} {\bibfnamefont
  {P.}~\bibnamefont {Salek}},\ }\href {https://doi.org/10.1063/1.2956507}
  {\bibfield  {journal} {\bibinfo  {journal} {The Journal of Chemical Physics}\
  }\textbf {\bibinfo {volume} {129}},\ \bibinfo {pages} {104101} (\bibinfo
  {year} {2008})}\BibitemShut {NoStop}%
\bibitem [{\citenamefont {Maier}, \citenamefont {Ikabata},\ and\ \citenamefont
  {Nakai}(2019)}]{doi:10.1021/acs.jctc.9b00228}%
  \BibitemOpen
  \bibfield  {author} {\bibinfo {author} {\bibfnamefont {T.~M.}\ \bibnamefont
  {Maier}}, \bibinfo {author} {\bibfnamefont {Y.}~\bibnamefont {Ikabata}},\
  and\ \bibinfo {author} {\bibfnamefont {H.}~\bibnamefont {Nakai}},\ }\href
  {https://doi.org/10.1021/acs.jctc.9b00228} {\bibfield  {journal} {\bibinfo
  {journal} {Journal of Chemical Theory and Computation}\ }\textbf {\bibinfo
  {volume} {15}},\ \bibinfo {pages} {4745} (\bibinfo {year} {2019})},\ \bibinfo
  {note} {pMID: 31403794}\BibitemShut {NoStop}%
\bibitem [{\citenamefont {Held}, \citenamefont {Hanrath},\ and\ \citenamefont
  {Dolg}(2018)}]{doi:10.1021/acs.jctc.8b00358}%
  \BibitemOpen
  \bibfield  {author} {\bibinfo {author} {\bibfnamefont {J.}~\bibnamefont
  {Held}}, \bibinfo {author} {\bibfnamefont {M.}~\bibnamefont {Hanrath}},\ and\
  \bibinfo {author} {\bibfnamefont {M.}~\bibnamefont {Dolg}},\ }\href
  {https://doi.org/10.1021/acs.jctc.8b00358} {\bibfield  {journal} {\bibinfo
  {journal} {Journal of Chemical Theory and Computation}\ }\textbf {\bibinfo
  {volume} {14}},\ \bibinfo {pages} {6197} (\bibinfo {year} {2018})},\ \bibinfo
  {note} {pMID: 30365307}\BibitemShut {NoStop}%
\bibitem [{\citenamefont {Schwegler}, \citenamefont {Challacombe},\ and\
  \citenamefont {Head-Gordon}(1997)}]{10.1063/1.473833}%
  \BibitemOpen
  \bibfield  {author} {\bibinfo {author} {\bibfnamefont {E.}~\bibnamefont
  {Schwegler}}, \bibinfo {author} {\bibfnamefont {M.}~\bibnamefont
  {Challacombe}},\ and\ \bibinfo {author} {\bibfnamefont {M.}~\bibnamefont
  {Head-Gordon}},\ }\href {https://doi.org/10.1063/1.473833} {\bibfield
  {journal} {\bibinfo  {journal} {The Journal of Chemical Physics}\ }\textbf
  {\bibinfo {volume} {106}},\ \bibinfo {pages} {9708} (\bibinfo {year}
  {1997})}\BibitemShut {NoStop}%
\bibitem [{\citenamefont {Rudberg}\ and\ \citenamefont
  {Sałek}(2006)}]{10.1063/1.2244565}%
  \BibitemOpen
  \bibfield  {author} {\bibinfo {author} {\bibfnamefont {E.}~\bibnamefont
  {Rudberg}}\ and\ \bibinfo {author} {\bibfnamefont {P.}~\bibnamefont
  {Sałek}},\ }\href {https://doi.org/10.1063/1.2244565} {\bibfield  {journal}
  {\bibinfo  {journal} {The Journal of Chemical Physics}\ }\textbf {\bibinfo
  {volume} {125}},\ \bibinfo {pages} {084106} (\bibinfo {year}
  {2006})}\BibitemShut {NoStop}%
\bibitem [{\citenamefont {Pulay}(2014)}]{https://doi.org/10.1002/wcms.1171}%
  \BibitemOpen
  \bibfield  {author} {\bibinfo {author} {\bibfnamefont {P.}~\bibnamefont
  {Pulay}},\ }\href {https://doi.org/10.1002/wcms.1171} {\bibfield  {journal}
  {\bibinfo  {journal} {WIREs Computational Molecular Science}\ }\textbf
  {\bibinfo {volume} {4}},\ \bibinfo {pages} {169} (\bibinfo {year}
  {2014})}\BibitemShut {NoStop}%
\bibitem [{\citenamefont {Yamaguchi}\ and\ \citenamefont
  {Schaefer~III}(2011)}]{doi:https://doi.org/10.1002/9780470749593.hrs006}%
  \BibitemOpen
  \bibfield  {author} {\bibinfo {author} {\bibfnamefont {Y.}~\bibnamefont
  {Yamaguchi}}\ and\ \bibinfo {author} {\bibfnamefont {H.~F.}\ \bibnamefont
  {Schaefer~III}},\ }\enquote {\bibinfo {title} {Analytic derivative methods in
  molecular electronic structure theory: A new dimension to quantum chemistry
  and its applications to spectroscopy},}\ in\ \href
  {https://doi.org/10.1002/9780470749593.hrs006} {\emph {\bibinfo {booktitle}
  {Handbook of High‐resolution Spectroscopy}}}\ (\bibinfo  {publisher} {John
  Wiley \& Sons, Ltd},\ \bibinfo {year} {2011})\BibitemShut {NoStop}%
\bibitem [{\citenamefont {Schweizer}, \citenamefont {Doser},\ and\
  \citenamefont {Ochsenfeld}(2008)}]{10.1063/1.2906127}%
  \BibitemOpen
  \bibfield  {author} {\bibinfo {author} {\bibfnamefont {S.}~\bibnamefont
  {Schweizer}}, \bibinfo {author} {\bibfnamefont {B.}~\bibnamefont {Doser}},\
  and\ \bibinfo {author} {\bibfnamefont {C.}~\bibnamefont {Ochsenfeld}},\
  }\href {https://doi.org/10.1063/1.2906127} {\bibfield  {journal} {\bibinfo
  {journal} {The Journal of Chemical Physics}\ }\textbf {\bibinfo {volume}
  {128}},\ \bibinfo {pages} {154101} (\bibinfo {year} {2008})}\BibitemShut
  {NoStop}%
\bibitem [{\citenamefont {Csóka}\ and\ \citenamefont
  {Kállay}(2023)}]{10.1063/5.0131683}%
  \BibitemOpen
  \bibfield  {author} {\bibinfo {author} {\bibfnamefont {J.}~\bibnamefont
  {Csóka}}\ and\ \bibinfo {author} {\bibfnamefont {M.}~\bibnamefont
  {Kállay}},\ }\href {https://doi.org/10.1063/5.0131683} {\bibfield  {journal}
  {\bibinfo  {journal} {The Journal of Chemical Physics}\ }\textbf {\bibinfo
  {volume} {158}},\ \bibinfo {pages} {024110} (\bibinfo {year}
  {2023})}\BibitemShut {NoStop}%
\bibitem [{\citenamefont {Frame}\ \emph {et~al.}(2018)\citenamefont {Frame},
  \citenamefont {He}, \citenamefont {Ipsen}, \citenamefont {Lee}, \citenamefont
  {Lee},\ and\ \citenamefont {Rrapaj}}]{PhysRevLett.121.032501}%
  \BibitemOpen
  \bibfield  {author} {\bibinfo {author} {\bibfnamefont {D.}~\bibnamefont
  {Frame}}, \bibinfo {author} {\bibfnamefont {R.}~\bibnamefont {He}}, \bibinfo
  {author} {\bibfnamefont {I.}~\bibnamefont {Ipsen}}, \bibinfo {author}
  {\bibfnamefont {D.}~\bibnamefont {Lee}}, \bibinfo {author} {\bibfnamefont
  {D.}~\bibnamefont {Lee}},\ and\ \bibinfo {author} {\bibfnamefont
  {E.}~\bibnamefont {Rrapaj}},\ }\href
  {https://doi.org/10.1103/PhysRevLett.121.032501} {\bibfield  {journal}
  {\bibinfo  {journal} {Phys. Rev. Lett.}\ }\textbf {\bibinfo {volume} {121}},\
  \bibinfo {pages} {032501} (\bibinfo {year} {2018})},\ \Eprint
  {https://arxiv.org/abs/1711.07090} {1711.07090} \BibitemShut {NoStop}%
\bibitem [{\citenamefont {Zhai}\ \emph {et~al.}(2023)\citenamefont {Zhai},
  \citenamefont {Larsson}, \citenamefont {Lee}, \citenamefont {Cui},
  \citenamefont {Zhu}, \citenamefont {Sun}, \citenamefont {Peng}, \citenamefont
  {Peng}, \citenamefont {Liao}, \citenamefont {Tölle}, \citenamefont {Yang},
  \citenamefont {Li},\ and\ \citenamefont {Chan}}]{Block2}%
  \BibitemOpen
  \bibfield  {author} {\bibinfo {author} {\bibfnamefont {H.}~\bibnamefont
  {Zhai}}, \bibinfo {author} {\bibfnamefont {H.~R.}\ \bibnamefont {Larsson}},
  \bibinfo {author} {\bibfnamefont {S.}~\bibnamefont {Lee}}, \bibinfo {author}
  {\bibfnamefont {Z.-H.}\ \bibnamefont {Cui}}, \bibinfo {author} {\bibfnamefont
  {T.}~\bibnamefont {Zhu}}, \bibinfo {author} {\bibfnamefont {C.}~\bibnamefont
  {Sun}}, \bibinfo {author} {\bibfnamefont {L.}~\bibnamefont {Peng}}, \bibinfo
  {author} {\bibfnamefont {R.}~\bibnamefont {Peng}}, \bibinfo {author}
  {\bibfnamefont {K.}~\bibnamefont {Liao}}, \bibinfo {author} {\bibfnamefont
  {J.}~\bibnamefont {Tölle}}, \bibinfo {author} {\bibfnamefont
  {J.}~\bibnamefont {Yang}}, \bibinfo {author} {\bibfnamefont {S.}~\bibnamefont
  {Li}},\ and\ \bibinfo {author} {\bibfnamefont {G.~K.-L.}\ \bibnamefont
  {Chan}},\ }\href {https://doi.org/10.1063/5.0180424} {\bibfield  {journal}
  {\bibinfo  {journal} {The Journal of Chemical Physics}\ }\textbf {\bibinfo
  {volume} {159}},\ \bibinfo {pages} {234801} (\bibinfo {year}
  {2023})}\BibitemShut {NoStop}%
\bibitem [{\citenamefont {Crespo-Otero}\ and\ \citenamefont
  {Barbatti}(2018)}]{CrespoOtero18-chemicalreview}%
  \BibitemOpen
  \bibfield  {author} {\bibinfo {author} {\bibfnamefont {R.}~\bibnamefont
  {Crespo-Otero}}\ and\ \bibinfo {author} {\bibfnamefont {M.}~\bibnamefont
  {Barbatti}},\ }\href {https://doi.org/10.1021/acs.chemrev.7b00577} {\bibfield
   {journal} {\bibinfo  {journal} {Chemical Reviews}\ }\textbf {\bibinfo
  {volume} {118}},\ \bibinfo {pages} {7026} (\bibinfo {year} {2018})},\
  \bibinfo {note} {pMID: 29767966}\BibitemShut {NoStop}%
\bibitem [{\citenamefont {Tully}(1990)}]{Tully90-FSSH}%
  \BibitemOpen
  \bibfield  {author} {\bibinfo {author} {\bibfnamefont {J.~C.}\ \bibnamefont
  {Tully}},\ }\href {https://doi.org/10.1063/1.459170} {\bibfield  {journal}
  {\bibinfo  {journal} {The Journal of Chemical Physics}\ }\textbf {\bibinfo
  {volume} {93}},\ \bibinfo {pages} {1061} (\bibinfo {year}
  {1990})}\BibitemShut {NoStop}%
\bibitem [{\citenamefont {Barbatti}\ \emph {et~al.}(2014)\citenamefont
  {Barbatti}, \citenamefont {Ruckenbauer}, \citenamefont {Plasser},
  \citenamefont {Pittner}, \citenamefont {Granucci}, \citenamefont {Persico},\
  and\ \citenamefont {Lischka}}]{NewtonX14}%
  \BibitemOpen
  \bibfield  {author} {\bibinfo {author} {\bibfnamefont {M.}~\bibnamefont
  {Barbatti}}, \bibinfo {author} {\bibfnamefont {M.}~\bibnamefont
  {Ruckenbauer}}, \bibinfo {author} {\bibfnamefont {F.}~\bibnamefont
  {Plasser}}, \bibinfo {author} {\bibfnamefont {J.}~\bibnamefont {Pittner}},
  \bibinfo {author} {\bibfnamefont {G.}~\bibnamefont {Granucci}}, \bibinfo
  {author} {\bibfnamefont {M.}~\bibnamefont {Persico}},\ and\ \bibinfo {author}
  {\bibfnamefont {H.}~\bibnamefont {Lischka}},\ }\href
  {https://doi.org/10.1002/wcms.1158} {\bibfield  {journal} {\bibinfo
  {journal} {WIREs Computational Molecular Science}\ }\textbf {\bibinfo
  {volume} {4}},\ \bibinfo {pages} {26} (\bibinfo {year} {2014})}\BibitemShut
  {NoStop}%
\bibitem [{\citenamefont {Barbatti}\ \emph {et~al.}(2022)\citenamefont
  {Barbatti}, \citenamefont {Bondanza}, \citenamefont {Crespo-Otero},
  \citenamefont {Demoulin}, \citenamefont {Dral}, \citenamefont {Granucci},
  \citenamefont {Kossoski}, \citenamefont {Lischka}, \citenamefont {Mennucci},
  \citenamefont {Mukherjee}, \citenamefont {Pederzoli}, \citenamefont
  {Persico}, \citenamefont {Pinheiro~Jr}, \citenamefont {Pittner},
  \citenamefont {Plasser}, \citenamefont {Sangiogo~Gil},\ and\ \citenamefont
  {Stojanovic}}]{NewtonX22}%
  \BibitemOpen
  \bibfield  {author} {\bibinfo {author} {\bibfnamefont {M.}~\bibnamefont
  {Barbatti}}, \bibinfo {author} {\bibfnamefont {M.}~\bibnamefont {Bondanza}},
  \bibinfo {author} {\bibfnamefont {R.}~\bibnamefont {Crespo-Otero}}, \bibinfo
  {author} {\bibfnamefont {B.}~\bibnamefont {Demoulin}}, \bibinfo {author}
  {\bibfnamefont {P.~O.}\ \bibnamefont {Dral}}, \bibinfo {author}
  {\bibfnamefont {G.}~\bibnamefont {Granucci}}, \bibinfo {author}
  {\bibfnamefont {F.}~\bibnamefont {Kossoski}}, \bibinfo {author}
  {\bibfnamefont {H.}~\bibnamefont {Lischka}}, \bibinfo {author} {\bibfnamefont
  {B.}~\bibnamefont {Mennucci}}, \bibinfo {author} {\bibfnamefont
  {S.}~\bibnamefont {Mukherjee}}, \bibinfo {author} {\bibfnamefont
  {M.}~\bibnamefont {Pederzoli}}, \bibinfo {author} {\bibfnamefont
  {M.}~\bibnamefont {Persico}}, \bibinfo {author} {\bibfnamefont
  {M.}~\bibnamefont {Pinheiro~Jr}}, \bibinfo {author} {\bibfnamefont
  {J.}~\bibnamefont {Pittner}}, \bibinfo {author} {\bibfnamefont
  {F.}~\bibnamefont {Plasser}}, \bibinfo {author} {\bibfnamefont
  {E.}~\bibnamefont {Sangiogo~Gil}},\ and\ \bibinfo {author} {\bibfnamefont
  {L.}~\bibnamefont {Stojanovic}},\ }\href
  {https://doi.org/10.1021/acs.jctc.2c00804} {\bibfield  {journal} {\bibinfo
  {journal} {Journal of Chemical Theory and Computation}\ }\textbf {\bibinfo
  {volume} {18}},\ \bibinfo {pages} {6851} (\bibinfo {year} {2022})},\ \bibinfo
  {note} {pMID: 36194696}\BibitemShut {NoStop}%
\bibitem [{\citenamefont {Butcher}(1965)}]{Butcher65-multistepODEintegrator}%
  \BibitemOpen
  \bibfield  {author} {\bibinfo {author} {\bibfnamefont {J.~C.}\ \bibnamefont
  {Butcher}},\ }\href {https://doi.org/10.1145/321250.321261} {\bibfield
  {journal} {\bibinfo  {journal} {J. ACM}\ }\textbf {\bibinfo {volume} {12}},\
  \bibinfo {pages} {124–135} (\bibinfo {year} {1965})}\BibitemShut {NoStop}%
\bibitem [{\citenamefont {Granucci}\ and\ \citenamefont
  {Persico}(2007)}]{Granucci07-decoherence}%
  \BibitemOpen
  \bibfield  {author} {\bibinfo {author} {\bibfnamefont {G.}~\bibnamefont
  {Granucci}}\ and\ \bibinfo {author} {\bibfnamefont {M.}~\bibnamefont
  {Persico}},\ }\href {https://doi.org/10.1063/1.2715585} {\bibfield  {journal}
  {\bibinfo  {journal} {The Journal of Chemical Physics}\ }\textbf {\bibinfo
  {volume} {126}},\ \bibinfo {pages} {134114} (\bibinfo {year}
  {2007})}\BibitemShut {NoStop}%
\bibitem [{\citenamefont {Head-Gordon}, \citenamefont {Maslen},\ and\
  \citenamefont {White}(1998)}]{10.1063/1.475423}%
  \BibitemOpen
  \bibfield  {author} {\bibinfo {author} {\bibfnamefont {M.}~\bibnamefont
  {Head-Gordon}}, \bibinfo {author} {\bibfnamefont {P.~E.}\ \bibnamefont
  {Maslen}},\ and\ \bibinfo {author} {\bibfnamefont {C.~A.}\ \bibnamefont
  {White}},\ }\href {https://doi.org/10.1063/1.475423} {\bibfield  {journal}
  {\bibinfo  {journal} {The Journal of Chemical Physics}\ }\textbf {\bibinfo
  {volume} {108}},\ \bibinfo {pages} {616} (\bibinfo {year}
  {1998})}\BibitemShut {NoStop}%
\bibitem [{\citenamefont {Amos}\ and\ \citenamefont
  {Hall}(1961)}]{10.1098/rspa.1961.0175}%
  \BibitemOpen
  \bibfield  {author} {\bibinfo {author} {\bibfnamefont {A.~T.}\ \bibnamefont
  {Amos}}\ and\ \bibinfo {author} {\bibfnamefont {G.~G.}\ \bibnamefont
  {Hall}},\ }\href {https://doi.org/10.1098/rspa.1961.0175} {\bibfield
  {journal} {\bibinfo  {journal} {Proc. R. Soc. A.}\ }\textbf {\bibinfo
  {volume} {263}},\ \bibinfo {pages} {483} (\bibinfo {year}
  {1961})}\BibitemShut {NoStop}%
\bibitem [{\citenamefont {King}\ \emph {et~al.}(1967)\citenamefont {King},
  \citenamefont {Stanton}, \citenamefont {Kim}, \citenamefont {Wyatt},\ and\
  \citenamefont {Parr}}]{10.1063/1.1712221}%
  \BibitemOpen
  \bibfield  {author} {\bibinfo {author} {\bibfnamefont {H.~F.}\ \bibnamefont
  {King}}, \bibinfo {author} {\bibfnamefont {R.~E.}\ \bibnamefont {Stanton}},
  \bibinfo {author} {\bibfnamefont {H.}~\bibnamefont {Kim}}, \bibinfo {author}
  {\bibfnamefont {R.~E.}\ \bibnamefont {Wyatt}},\ and\ \bibinfo {author}
  {\bibfnamefont {R.~G.}\ \bibnamefont {Parr}},\ }\href
  {https://doi.org/10.1063/1.1712221} {\bibfield  {journal} {\bibinfo
  {journal} {J. Chem. Phys.}\ }\textbf {\bibinfo {volume} {47}},\ \bibinfo
  {pages} {1936} (\bibinfo {year} {1967})}\BibitemShut {NoStop}%
\bibitem [{\citenamefont {Chen}\ and\ \citenamefont
  {Scuseria}(2023)}]{10.1063/5.0156124}%
  \BibitemOpen
  \bibfield  {author} {\bibinfo {author} {\bibfnamefont {G.~P.}\ \bibnamefont
  {Chen}}\ and\ \bibinfo {author} {\bibfnamefont {G.~E.}\ \bibnamefont
  {Scuseria}},\ }\href {https://doi.org/10.1063/5.0156124} {\bibfield
  {journal} {\bibinfo  {journal} {J. Chem. Phys.}\ }\textbf {\bibinfo {volume}
  {158}},\ \bibinfo {pages} {231102} (\bibinfo {year} {2023})}\BibitemShut
  {NoStop}%
\bibitem [{\citenamefont {Miller}\ and\ \citenamefont
  {Parker}(2025)}]{10.1063/5.0246790}%
  \BibitemOpen
  \bibfield  {author} {\bibinfo {author} {\bibfnamefont {E.~R.}\ \bibnamefont
  {Miller}}\ and\ \bibinfo {author} {\bibfnamefont {S.~M.}\ \bibnamefont
  {Parker}},\ }\href {https://doi.org/10.1063/5.0246790} {\bibfield  {journal}
  {\bibinfo  {journal} {J. Chem. Phys.}\ }\textbf {\bibinfo {volume} {162}},\
  \bibinfo {pages} {104115} (\bibinfo {year} {2025})}\BibitemShut {NoStop}%
\end{thebibliography}%
